\documentclass[a4paper,fleqn,usenatbib]{mnras}

\usepackage{newtxtext,newtxmath}

\usepackage[T1]{fontenc}
\usepackage{ae,aecompl}

\usepackage{graphicx}	
\usepackage{amsmath}	
\usepackage{booktabs}   
\usepackage{multirow}   
\usepackage{rotating}   %
\usepackage{natbib}
\usepackage{color, soul} 
\soulregister\cite7
\soulregister\citet7
\soulregister\citep7
\soulregister\ref7
\usepackage{xcolor} 
\usepackage{xspace}
\usepackage{subcaption}     
\usepackage[stable]{footmisc}
\usepackage{siunitx}
\usepackage{pdflscape}
\usepackage{xcolor}
\usepackage{etoolbox} 
\usepackage{placeins}
\usepackage{float}
\floatstyle{plaintop} 
\restylefloat{table}

\renewcommand{\textbf}{}  

\title[From detailed host stellar abundances to planetary interiors]{Detailed chemical compositions of planet-hosting stars: II. Exploration of the interiors of terrestrial-type exoplanets}  

\author[H.S.\ Wang et al.] 
{H. S.\ Wang$^{1,2 \thanks{\textrm{E-mail: haiwang@phys.ethz.ch}}}$,
	S. P.\ Quanz$^{1,2}$, 	
	D.\ Yong$^{3}$,
	F.\ Liu$^{4}$, 
	F.\ Seidler$^{1}$,
	L.\ Acu\~{n}a$^{5}$,
	and S. J. Mojzsis$^{6,7}$ \\
	\\ 
	$^{1}$Institute for Particle Physics and Astrophysics, ETH Z\"urich, Wolfgang-Pauli-Strasse 27, 8093 Z\"urich, Switzerland\\
	$^{2}$National Center of Competence in Research PlanetS (www.nccr-planets.ch) \\
	$^{3}$Research School of Astronomy and Astrophysics, Australian National University, Canberra, ACT 2611, Australia\\
	$^{4}$Centre for Astrophysics and Supercomputing, Swinburne University of Technology, Hawthorn, Victoria 3122, Australia\\
	$^{5}$Aix-Marseille Univ., CNRS, CNES, LAM, Marseille, France\\
	$^{6}$Origins Research Institute, Research Centre for Astronomy and Earth Sciences, H-1112 Budapest, Hungary \\
	$^{7}$Department of Geological Sciences, University of Colorado, Boulder, CO 80309-0399, USA \\
}
\date{Accepted 2022 April 19. Received 2022 April 15; in original form 2021 October 6}

\pubyear{2022}

\begin{document}
\label{firstpage}
\pagerange{\pageref{firstpage}--\pageref{lastpage}}
\maketitle

\begin{abstract}
A major goal in the discovery and characterisation of exoplanets is to identify terrestrial-type worlds that are similar to (or otherwise distinct from) our Earth. 
Recent results have highlighted the importance of applying devolatilisation -- i.e. depletion of volatiles -- to the chemical composition of planet-hosting stars to constrain bulk composition and interiors of terrestrial-type exoplanets. 
In this work, we apply such an approach to a selected sample of 13 planet-hosting Sun-like stars, for which high-precision photospheric abundances have been determined in the first paper of the series. 
With the resultant devolatilised stellar composition (i.e. the model planetary bulk composition) as well as other constraints including mass and radius, we model the detailed mineralogy and interior structure of hypothetical, habitable-zone terrestrial planets ("exo-Earths") around these stars. Model output shows that most of these exo-Earths are expected to have broadly Earth-like composition and interior structure, consistent with conclusions derived independently from analysis of polluted white dwarfs. 
The exceptions are the Kepler-10 and Kepler-37 exo-Earths, which we predict are strongly oxidised and thus would develop metallic cores much smaller than Earth.  
Investigating our devolatilisation model at its extremes as well as varying planetary mass and radius (within the terrestrial regime) reveals potential diversities in the interiors of terrestrial planets. 
By considering (i) high-precision stellar abundances, (ii) devolatilisation, and (iii) planetary mass and radius holistically, this work represents essential steps to explore the detailed mineralogy and interior structure of terrestrial-type exoplanets, which in turn are fundamental for our understanding of planetary dynamics and long-term evolution. 
\end{abstract}




\section{INTRODUCTION}
\label{sec:intro}
An important threshold has been crossed for detailed studies of interior structure and composition of terrestrial exoplanets following the ever more precise measurements of both planetary mass and radius \citep{Weiss2016, Stassun2017, Stassun2018, Otegi2020} and of host stellar photospheres that reveal the primordial elemental compositions of the systems \citep{Nissen2015, Brewer2016, Liu2016, Liu2020, Delgado2017, Bedell2018, Clark2021, Adibekyan2021}. Indeed, the past decade has witnessed a number of interior models that follow this lead \citep[e.g.][]{Dorn2015, Santos2015, Unterborn2016, Dorn2017, Brugger2017, Unterborn2018, Wang2019b, Acuna2021, Wang2022}. We can anticipate that future observations of planetary atmospheres with JWST \citep{Morley2017, Gialluca2021} and other innovative ground- and space-based missions and mission concepts such as ELT/METIS \citep{Quanz2015, Bowens2021}, ARIEL \citep{Tinetti2018, Turrini2021}, PLATO \citep{Rauer2014, Nascimbeni2022}, and LIFE \citep{Quanz2021, Quanz2022} will reveal new details of the surface and interior characteristics of terrestrial exoplanets, given the evolutionary outcomes of the dynamic interactions between the interior, surface, atmosphere, and possible hydrosphere or even biosphere \citep{Shahar2019, Bower2019, Ortenzi2020, Dyck2021, Hakim2021, Acuna2021, Kacar2021}. 

In the present work, we build upon the procedures outlined in \cite{Wang2019a, Wang2019b} (W19a and b, thereafter) that introduced the idea of using the devolatilised host stellar abundances, rather than the unaltered host stellar abundances, to constrain the bulk composition and interior modelling of hypothetical habitable-zone terrestrial exoplanets ("exo-Earths"). This idea was established based on the observations of bulk composition differences and similarities of the Solar System's rocky bodies relative to the Sun \citep{Grossman1974, Bland2005, Davis2006, Carlson2014, Wang2018, Sossi2018}. Importantly, among the 10 major rock-forming elements (Mg, Si, Fe, Ni, Al, Ca, Na, O, S, and C), only Ca and Al (the two most refractory ones) are not observed to be depleted in rocky bodies relative to the Sun. All other elements have been depleted to some degree: e.g. for Mg, Si, Fe and Ni the depletion is by $\sim$ 10-20\%; for volatiles like O, S and C the depletion is over 80\% (W19a). It is reasonable to argue that the devolatilisation process is not unique to the Solar System and may be a universal process in the formation of rocky (exo)planets. A recent study of the major rock-forming elements (including oxygen) for a sample of six white dwarfs \citep{Doyle2019} shows that the bulk composition of the planetary debris polluting these white dwarfs resembles those of rocky planets in the Solar System. This suggests that the parent rocky bodies of these debris must also be the devolatilised pieces of their host stars \citep[also see][]{Harrison2021, Bonsor2021}, although the exact devolatilisation factors for such planetary systems are beyond what can be constrained with the existing data.  Hence, starting with the best-known calibration of the devolatilisation based on our Solar System (in particular, Sun and Earth; W19a), we can apply it, to first order, to other Sun-like star systems and estimate the potential rocky planetary bulk composition from the measurable host stellar photospheric abundances. Subsequently, planetary interior composition and structure can be modelled, as shown for a sample of 4 planet-hosting stars in W19b. Even further, interior dynamics and thermo-chemical evolution \citep{Spaargaren2020, Spaargaren2021, Wang2020}, carbon cycle modelling \citep{Hakim2021}, and habitability arguments \citep{Kacar2021} ensue from such an analysis.

Here, we extend the analysis to a further sample of 13 planet-hosting Sun-like stars for which detailed and precise chemical compositions (for up to 18 elements including all major rock-forming elements at a typical precision of $\sim$ 0.025 dex) have been determined with high-quality spectra in the first paper of the series \citep{Liu2020} (hereafter Paper I). We are interested in how diverse the interiors of the model exo-Earths around these stars would be. 

The paper is organised by presenting our methodology and analysis in Section \ref{sec:sample} and results in Section \ref{sec:results}, followed by a discussion of the effect of varying devolatilisation scaling factors (within plausible bounds) and planetary size (within the terrestrial regime) on the interiors and of the model restrictions in Section \ref{sec:disc}. We summarise and conclude in Section \ref{sec:conclusion}. 

\section{METHODOLOGY AND ANALYSIS}
\label{sec:sample}
\subsection{Methodology}
To carry out the analysis, two sets of software are employed: \textit{ExoInt} (W19b) for devolatilising stellar abundances and modelling stoichiometric mantle and core compositions as well as core mass fractions; \textit{Perple\_X} \citep{Connolly2009} for modelling of a detailed mantle mineralogy and interior structure (e.g. self-consistent density, pressure and temperature profiles as well as core radius fraction). It is important to note that the devolatilisation model of W19a that we adopted is empirically quantified by the bulk elemental abundance ratio ($f$) between Earth and proto-Sun as a function of 50\% condensation temperature \citep[$T_c$;][]{Lodders2003}: 
\begin{equation}
\log(f) = \alpha\log(T_c) + \beta
\end{equation} 
where, the best-fit coefficients $\alpha=3.676 \pm 0.142$ and $\beta=-11.556 \pm 0.436$. 

An application of such a model to other planetary systems is by all means a simplification of devolatilisation processes and outcomes that may vary across different systems and, in principle, even at different orbital distances within one system. We therefore limit such an application to only habitable-zone, terrestrial-type exoplanets around Sun-like stars -- i.e. exo-Earths by our definition. It has been shown in the literature \citep[e.g.][]{Wang2018b, Sossi2018, Yoshizaki2020} that the bulk compositional differences of Venus and Mars from the Sun are within the uncertainty of such a difference between Earth and Sun. We emphasise that the empirical devolatilisation model of W19a is the first-order quantitative model of such an important process. A sophisticated model of devolatilisation that involves disc evolution, accretion and hydrodynamic escape processes is still awaiting formalisation \citep[e.g.][]{Wang2022b}. To account for potential variation in devolatilisation scales for the exo-Earths considered here, we vary the uncertainty range of the adopted W19a model arbitrarily by a factor of 3 to assess how this may affect the interior modelling results (to be discussed in Sect. \ref{sec:variance}).

For the detailed procedure of \textit{ExoInt}, we refer the reader to Fig. 3 and Appendix A of W19b. It is briefly summarised here that we adopt the chemical networks of Na$_2$O-CaO-MgO-Al$_2$O$_3$-SiO$_2$-FeO-NiO-SO$_3$-CO$_2$-C(graphite/diamond)-metals for the mantle composition (in terms of first-order oxides and reduced phases) and of Fe-Ni-Si-S alloy for the core composition of a terrestrial-type exoplanet. The core mass fraction is determined by mass balance after distributing the bulk planetary composition into the stoichiometric mantle and core compositions. The difference to W19b is that Si is not firstly oxidised in the oxidation sequence of major elements, but follows Na, Ca, Mg, and Al and precedes Fe, Ni, and S. This approach allows for the possibility that Si can be partially oxidised in the case of a reduced mantle (i.e. at a low oxidation state) and complies with the fact that it can be an important light element constituent of metallic cores \citep{McDonough2003, Hirose2013, Li2014, Wang2018}. This update has been applied in \cite{Wang2022} and the code is publicly accessible\footnote{\url{https://github.com/astro-seanwhy/ExoInt/tree/master/v1.2} (IDL version); \url{https://github.com/astro-seanwhy/ExoInt/tree/master/pyExoInt} (Python version).}.  

Having determined stoichiometric mantle and core compositions as well as core mass fractions, planetary mineralogy (i.e. complex mineral assemblages) and structure are then modeled with \textit{Perple\_X} by assuming that these exo-Earths are all Earth-like in mass and radius. Later in the discussion (Sec. \ref{sec:MR}), we explore the effect of varied planetary mass and radius (within the terrestrial density regime) on the predictions of detailed interiors.  
The underlying method for computing mantle mineralogy given the mantle composition of major oxides and the pressure and temperature profiles for a terrestrial-type planet is Gibbs free energy minimisation \citep{Connolly2009}. The mineral equations of state and thermodynamic parameters -- essential for the Gibbs free energy minimisation -- are adopted from \cite{Stixrude2011}. For a given Fe-Ni-Si-S-alloy core, we adopt the equation of state from \cite{Kuwayama2020}. We adopt an adiabatic thermal gradient, as similarly practiced in \cite{Dorn2015}, \cite{Unterborn2018}, and \cite{Lorenzo2018}, and integrate it with a mantle potential temperature of 1700 K at 1 bar, arbitrarily set to be approximate to that of the modern Earth \citep{Anderson2000}. \cite{Hinkel2018} found that different setups of a mantle potential temperature (at a typical range of 1500 K and 1900 K) for a terrestrial-type planet only introduce accountable effect towards the mineralogies in the transition zone between the upper and lower mantle -- we have verified this finding with Earth as an example (Fig. \ref{fig:mineral_Earth_varyTp.pdf}). 
We also introduce a temperature jump at the core mantle boundary (CMB) by following \cite{Noack2020} (for all of these exo-Earths) and \cite{Stixrude2014} (for the tested cases with masses out of $[0.8, 2] M_\oplus$ -- the range in which the \cite{Noack2020} model is parameterised). It has been found that a variable thermal profile plays a negligible role in changing the interior structure \citep{Dorn2015} -- we have further verified this finding with varied temperature jumps at the CMB in the case of Earth (Fig. \ref{fig:Various_Delta_T_CMB.pdf}). 

\subsection{Sample selection and analysis}
\begin{figure}
	\includegraphics[trim=0cm 5cm 1cm 6cm, scale=0.42,angle=0]{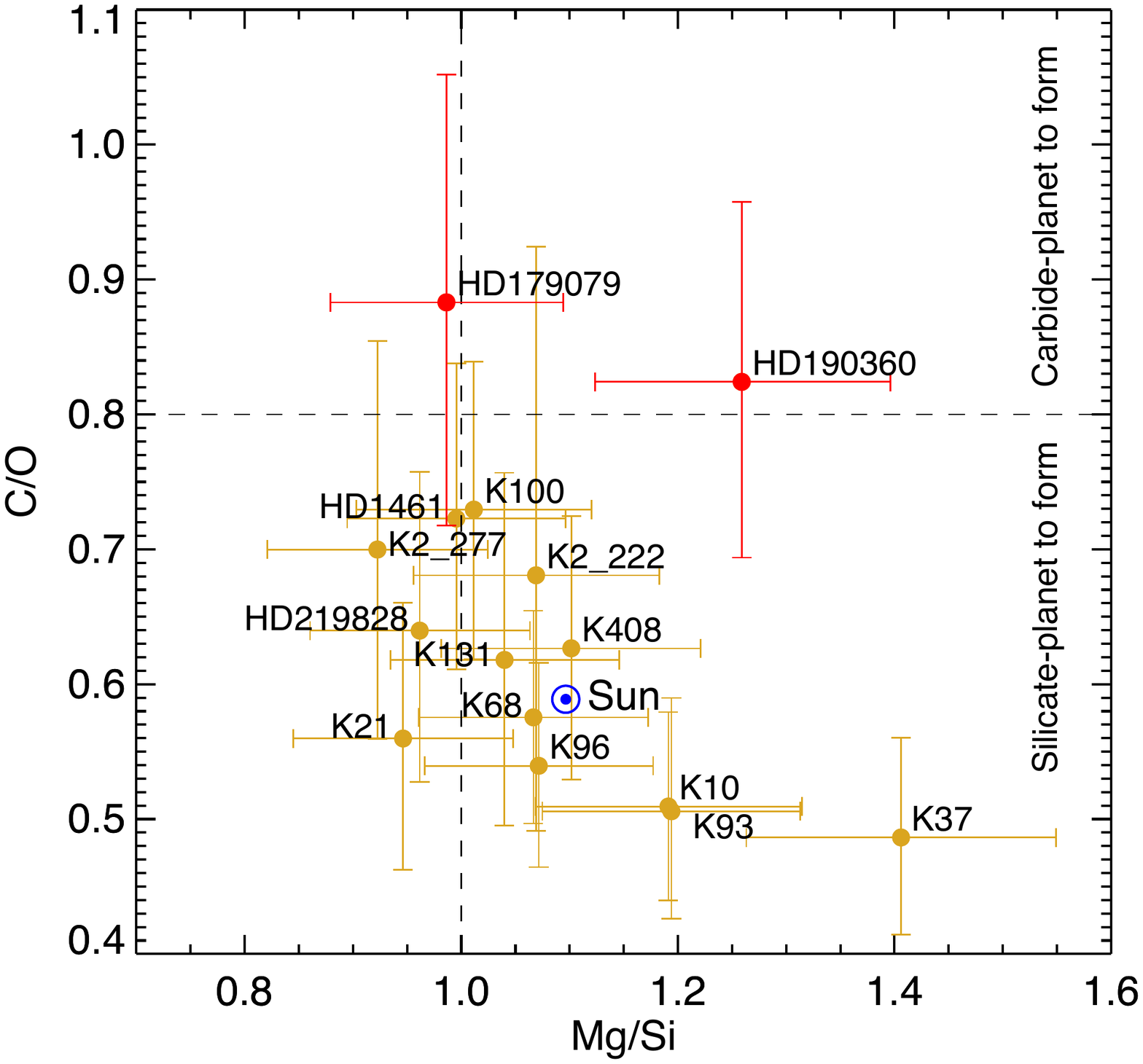} 
	\caption{C/O vs. Mg/Si -- abundance ratios by number -- for the sample of stars. The vertical and horizontal dashed lines respectively indicate the critical value of Mg/Si (= 1.0) and C/O (= 0.8) \citep{Bond2010, Suarez-Andres2018}. HD179079 and HD190360 (colored in red), for which C/O > 0.8, are excluded from our detailed analysis. The Sun \citep[`$\sun$';][]{Asplund2021} is shown for reference.}   
	\label{fig:star_ratios}
\end{figure}

Our sample for this study is based on the planet-hosting stars studied in Paper I: Kepler-21, Kepler-37, Kepler-68, Kepler-93, Kepler-96, Kepler-K100, Kepler-131, K2-222 (EPIC220709978), K2-277 (EPIC212357477), Kepler-408, HD1461, and HD219828. The addition to this list is Kepler-10, for which the detailed elemental abundances are obtained in \cite{Liu2016} in the same fashion as in Paper I. We have excluded Kepler-409 because its oxygen abundance -- essential for our analysis -- is undetermined.  
We have also excluded HD179079 and HD190360, for which the values of C/O are larger than 0.8 (Fig. \ref{fig:star_ratios}) and thus around which carbide planets may be developed \citep{Bond2010, Teske2014, Brewer2016}. Our model is based on silicate chemistry and thus does not apply to plausible carbide planets. For the reader interested in the interiors of carbide planets we refer to \cite{Hakim2018, Hakim2019} for more details. Therefore, we have a sample of 13 Sun-like stars (with $5400 < T_\textrm{eff} < 6400$ K, $4.0 < \log g < 4.5$ $\si{cm.s^{-2}}$, and $-0.3 < \textrm{[Fe/H]} < +0.3$), each of which has been confirmed to host at least one planet with a mass $<10 M_\oplus$ (except for HD 219828 that hosts a planet with a minimum mass of $\sim 21 M_\oplus$). Since these detected planets are all in the proximity (with an orbital period $< 50$ days) of their host stars, they are not included in this study that focuses on hypothetical, (habitable-zone) exo-Earths.

As shown in Fig. \ref{fig:star_ratios}, the values of Mg/Si (abundance ratio by number) in our sample are distributed within a narrow range from $\sim$ 0.9 to $\sim$ 1.4. Mg/Si modulates the dominant mineral phases in the mantle of a silicate planet: pyroxene (MgSiO$_3$) and various feldspars for Mg/Si < 1, a mixture of olivine (Mg$_2$SiO$_4$) and pyroxene assemblages for 1 < Mg/Si <2, and olivine with other Mg-rich species for Mg/Si > 2 \citep{Bond2010, Suarez-Andres2018}. At first glance, therefore, the mantles of exo-Earths around our sample of stars would most likely be made of a mixture of olivine and pyroxene assemblages, while some (with Mg/Si < 1) may be slightly enriched in pyroxenes.  

Following the study of W19b, we apply the Sun-to-Earth devolatilisation model (W19a) to the sample of stars for 10 major rock-forming elements: Mg, Si, Fe, Ni, Al, Ca, Na, O, S, and C. The differential abundances determined in Paper I (their Table 2) are firstly converted to the absolute abundances by referring to the latest solar abundances of \cite{Asplund2021}. We ignore the diffusion effect since this effect is equivalent for those major rock-forming elements for the Sun \citep{Asplund2009} and presumably for Sun-like stars as well. We also do not consider the effect of Galactic chemical evolution (GCE) on the host stellar abundances, since planets are fundamentally correlated with the properties (and formation environment) of \textit{individual} host stars. In other words, whatever a GCE effect may have with the host stellar abundances should have also been an inherent part of the formation histories of planets around these stars. Thus, the GCE effect has \textit{validly} shaped the chemical compositions of these planets. The resultant, "devolatilised" stellar abundances -- i.e., the model planetary bulk composition -- are listed in Table \ref{tab:abu} and used as a principal set of constraints for a detailed modelling of the interiors of (hypothetical) exo-Earths around these stars.

\section{RESULTS}
\label{sec:results}

\subsection{Key planetary geochemical ratios}
\label{sec:ratios}
The abundance ratio of carbon to oxygen (hereafter C/O) in a planet host star is useful, as illustrated in Fig. \ref{fig:star_ratios}, to indicate to first order if a potential rocky world around it would be dominated by silicates or by carbides. However, upon the application of the devolatilisation, both oxygen and carbon become so severely depleted that the remaining atoms are principally locked in planetary mineral assemblages. C/O in a rocky planet is therefore no longer a valid indicator of the mantle oxidation state, which, however, is essential to understand the planetary interiors. 

For a silicate planet including our own, MgO and SiO$_2$ are the foremost mineral oxides in the mantle, with Fe being distributed between its oxidised form (e.g. FeO) in the mantle and its reduced, metallic/liquid iron phase in the core, depending on the oxygen fugacity ($f\textrm{O}_2$) of the planet \citep[][]{McDonough1995, Palme2014b, Dorn2015, Wang2019b}. The calculation of $f\textrm{O}_2$, often relative to either the quartz-faylite-magnetite buffer \citep{Oneill1987, Cottrell2011} or the iron-w\"{u}stite buffer \citep[e.g.][]{Doyle2019}, requires the prior knowledge of the relative fractions between different phases of iron, which we do not have in the first place for exoplanets. Based on the estimated bulk elemental composition (Table \ref{tab:abu}) and considering that O, Mg, Si, and Fe are the foremost abundant, rock-forming elements \citep{Palme2014b, Wang2018}, we propose the bulk (O-Mg-2Si)/Fe as a simple alternative of oxygen fugacity to indicate the oxidation state of a silicate terrestrial exoplanet. 

Planetary Mg/Si (having no significant difference from its host stellar Mg/Si) is still critical to modulate the dominant mineral assemblages (olivine vs. pyroxene) in the mantle of a silicate planet. Further, Fe/Mg is preferred over Fe/Si as an indicator of the degree of core-mantle fractionation (determining the core size), owing to the fact that Si may be present as a major light element in the core of a rocky planet \citep{McDonough2003, Hirose2013, Li2014, Wang2018} and is indeed considered in the core compositional model in this work. 

\begin{figure*}
	\includegraphics[trim=3.0cm 2.0cm 3.0cm 2cm, scale=0.65,angle=90]{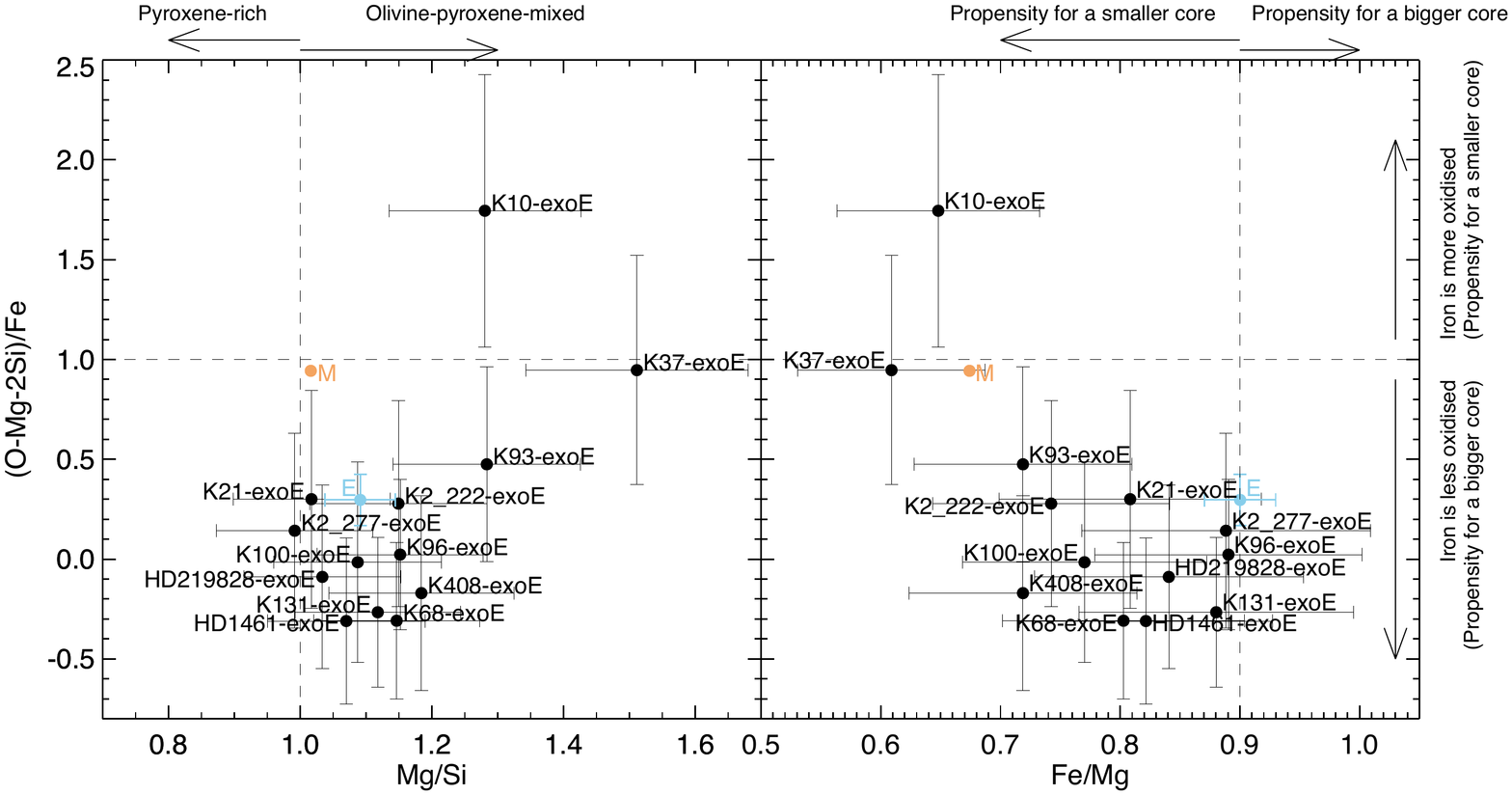} 
	\caption{(O - Mg - 2Si)/Fe vs. Mg/Si and (O - Mg - 2Si)/Fe vs. Fe/Mg diagrams for the sample of model exo-Earths. The dashed, vertical and horizontal lines classify these planets into different categories, in which different planetary interior properties may be expected (see text as well as the brief notations on the plot). Earth \citep["E", the blue dot;][]{Wang2018} and Mars \citep["M", the brown dot;][]{Yoshizaki2020} are shown for reference.}
	\label{fig:keyR}
\end{figure*}
Fig. \ref{fig:keyR} shows the distributions of these model exo-Earths on the diagram of (O - Mg - 2Si)/Fe vs. Mg/Si and (O - Mg - 2Si)/Fe vs. Fe/Mg. First, it shows that all of the sample planets are significantly below the unit line of (O - Mg - 2Si)/Fe, except Kepler-10 exo-Earth ("K10-exoE") and Kepler-37 exo-Earth ("K37-exoE"). The direct implication is that most of these planets would have a large iron core and potentially an Earth-like structure while iron in K10-exoE and K37-exoE may be much more oxidised and thus these two planets would develop comparably smaller cores. The core size of K10-exoE would be the smallest due to its significantly high (O - Mg - 2Si)/Fe that would cause most of the iron to be oxidised and locked in the mantle. Considering that the core size is also modulated by Fe/Mg (right panel of Fig. \ref{fig:keyR}), K37-exoE -- which has the equivalent oxidation state as Mars, but lower Fe/Mg -- would potentially develop a core that is smaller, in relative terms, than the core:mantle ratio of Mars. \cite{Frank2014} referred to this class of terrestrial-type exoplanets as "Super-Lunas". Among those planets clumped around the Earth's loci on both panels of the diagram, their mantle mineralogies would be more or less the same, with K93-exoE being likely the most olivine-rich due to its relatively high Mg/Si. These qualitative analyses are further verified by the following, detailed interior modelling. 

\subsection{Mantle and core compositions and core mass fraction}
\label{sec:mantle}

\begin{figure*}
\includegraphics[trim=0cm 0cm 0cm 0cm, scale=0.65,angle=0]{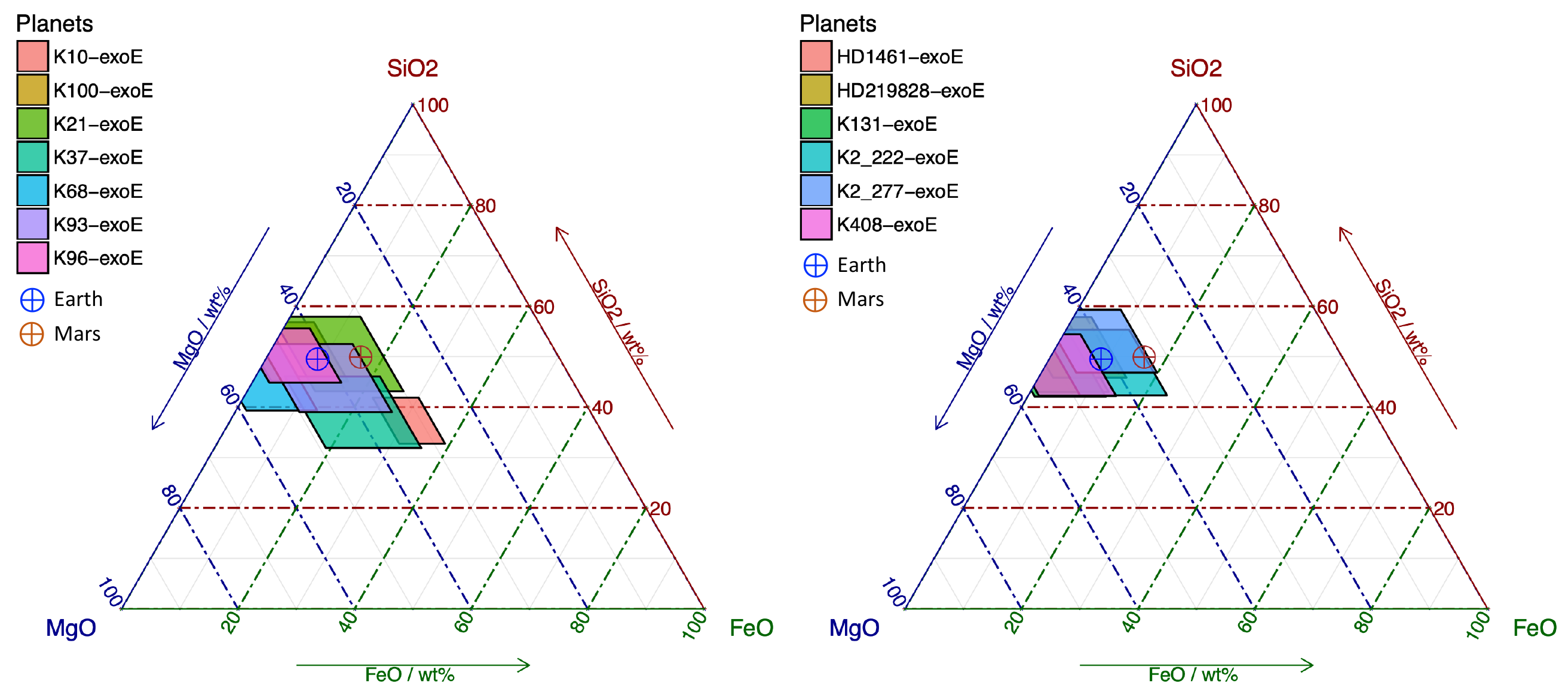}
\caption{Ternary diagrams showing the estimated mantle compositions of major mineral oxides (normalized by SiO2 + MgO + FeO = 100 wt\%) for the sample of model exo-Earths (for clarity, the sample is divided arbitrarily to two groups). The range of the squares indicates the modelled 1$\sigma$ uncertainties in the normalised mantle compositions of the individual planets. The normalised compositions of SiO2, MgO, and FeO of Earth mantle \citep{McDonough1995} and of Martian mantle \citep{Yoshizaki2020} are shown for reference.}
	\label{fig:ternary}
\end{figure*}

Our estimates of the mantle composition for all studied exo-Earths are presented in Table \ref{tab:interior}, with the normalised composition of the foremost major oxides (i.e. normalising the sum of SiO$_2$, MgO, and FeO to be 100 wt\%) shown in the ternary diagram Fig. \ref{fig:ternary}. Overall, the mantle compositions of these exo-Earths are very similar, except for K10-exoE and K37-exoE. The latter two are particularly enriched in FeO, consistent with the analysis above based on key geochemical ratios. 

\begin{figure*}
    \includegraphics[trim=1.0cm 1.0cm 2.0cm 0cm, scale=0.65,angle=90]{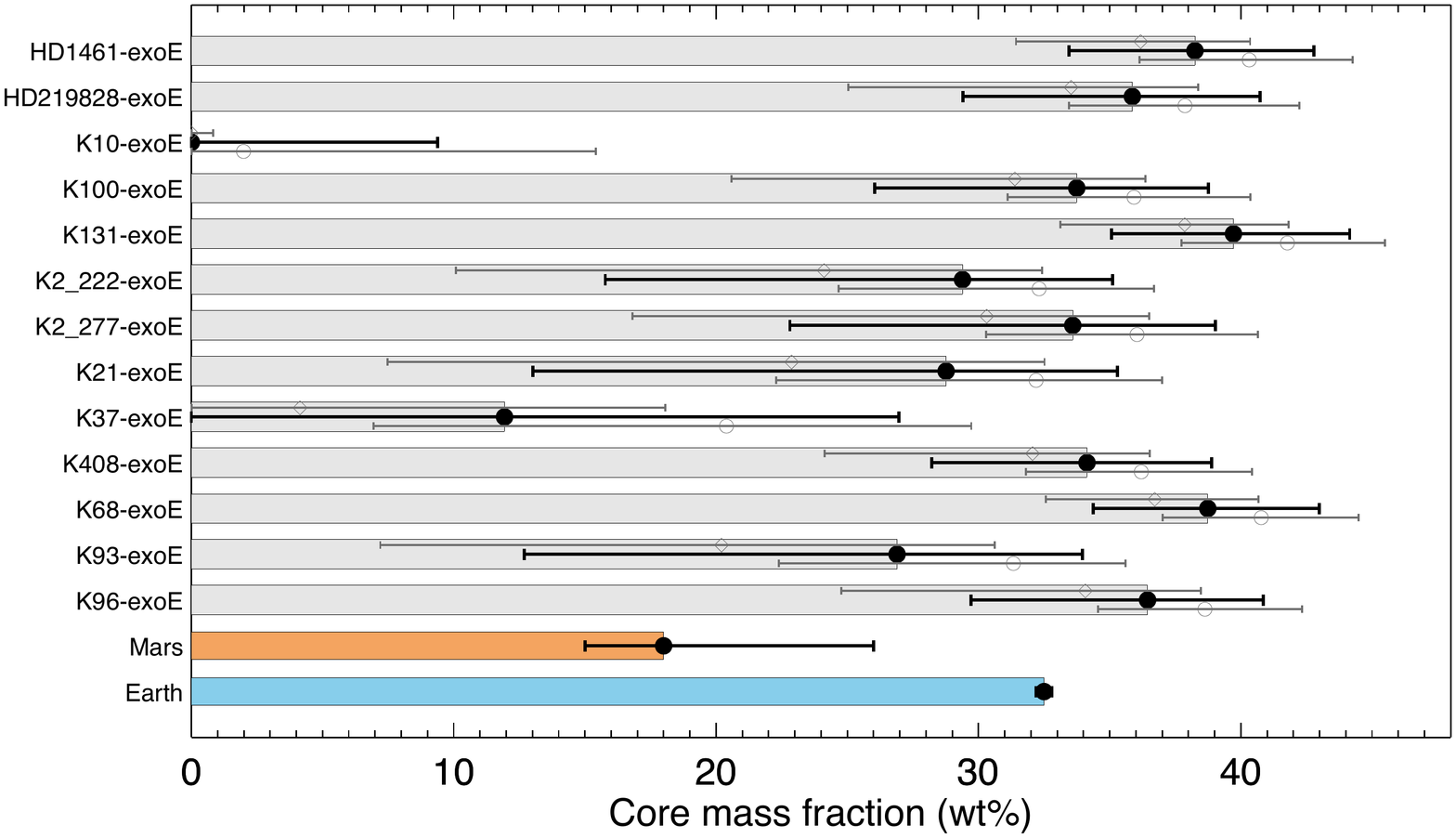} 
	\caption{Estimates of the core mass fraction (CMFs) of the sample of model exo-Earths under three different scenarios: (i) applying the standard Sun-to-Earth devolatilisation pattern (i.e., with the 1$\sigma$-uncertainty range) of W19a (filled circles with error bars in black); (ii) applying the 3$\sigma$ upper limit of the standard devolatilisation model (open diamonds with error bars in grey); (iii) applying the 3$\sigma$ lower limit of the standard devolatilisation model (open circles with error bars in grey). Earth's core mass fraction \citep[32.5 $\pm$ 0.3 wt\%;][]{Wang2018} and Mars' core mass fraction \citep[$18^{+8}_{-3}$ wt\%;][]{Yoshizaki2020} are shown for reference.} 
	\label{fig:cmf}
\end{figure*}

The estimates of core composition and core mass fraction (CMF; i.e. the core-to-planet mass ratio) of these exo-Earths are also presented in Table \ref{tab:interior}. We find that the core compositions of these exo-Earths are more or less the same, with the concentration of Fe ranging from $\sim$ 80 to $\sim$90 wt\%. In contrast, the values of CMF are diverse, ranging from $\sim$ 0 wt\% to $\sim$ 40 wt\% (Fig. \ref{fig:cmf}). K10-exoE and K37-exoE, for which the concentrations of FeO in the mantle are the highest (as mentioned above), are the ones with the lowest CMFs -- $0.0^{+9.4}_{-0.0}$ and $11.9^{+15.0}_{-11.9}$ - broadly mimicking Moon-like and Mars-like structures, respectively. The CMFs of eight exo-Earths (K21-exoE, K93-exoE, K96-exoE, K100-exoE, K408-exoE, K2-222-exoE, K2-227-exoE, and HD219828) are consistent with the CMF ($32.5\pm0.3$ wt\%; \citealt{Wang2018}) of the Earth within uncertainties. The remaining three exo-Earths (K68-exoE, K131-exoE, and HD 1461-exoE) have CMFs statistically higher than that of the Earth, with K131-exoE -- appearing at the bottom right (i.e. with the highest Fe/Mg and lowest (O-Mg-2Si)/Fe) in the right panel of Fig. \ref{fig:keyR} -- being the highest ($39.7^{+4.4}_{-4.7}$ wt\%) . 

The minor oxides -- Na$_2$O, CaO, and Al$_2$O$_3$ -- will be combined with the aforementioned three major oxides to model the detailed mineralogy of these exo-Earths. Other minor/trace end-members including NiO, SO$_3$, CO$_2$, C and metals are however not involved in the subsequently detailed interior modelling but their realisation is essential to correctly distribute oxygen into those more abundant oxides. 

\subsection{Mineralogy and internal structure}
\label{sec:mineral}

\begin{figure*}
	\includegraphics[trim=0cm 0cm 0cm 0cm, scale=0.70,angle=0]{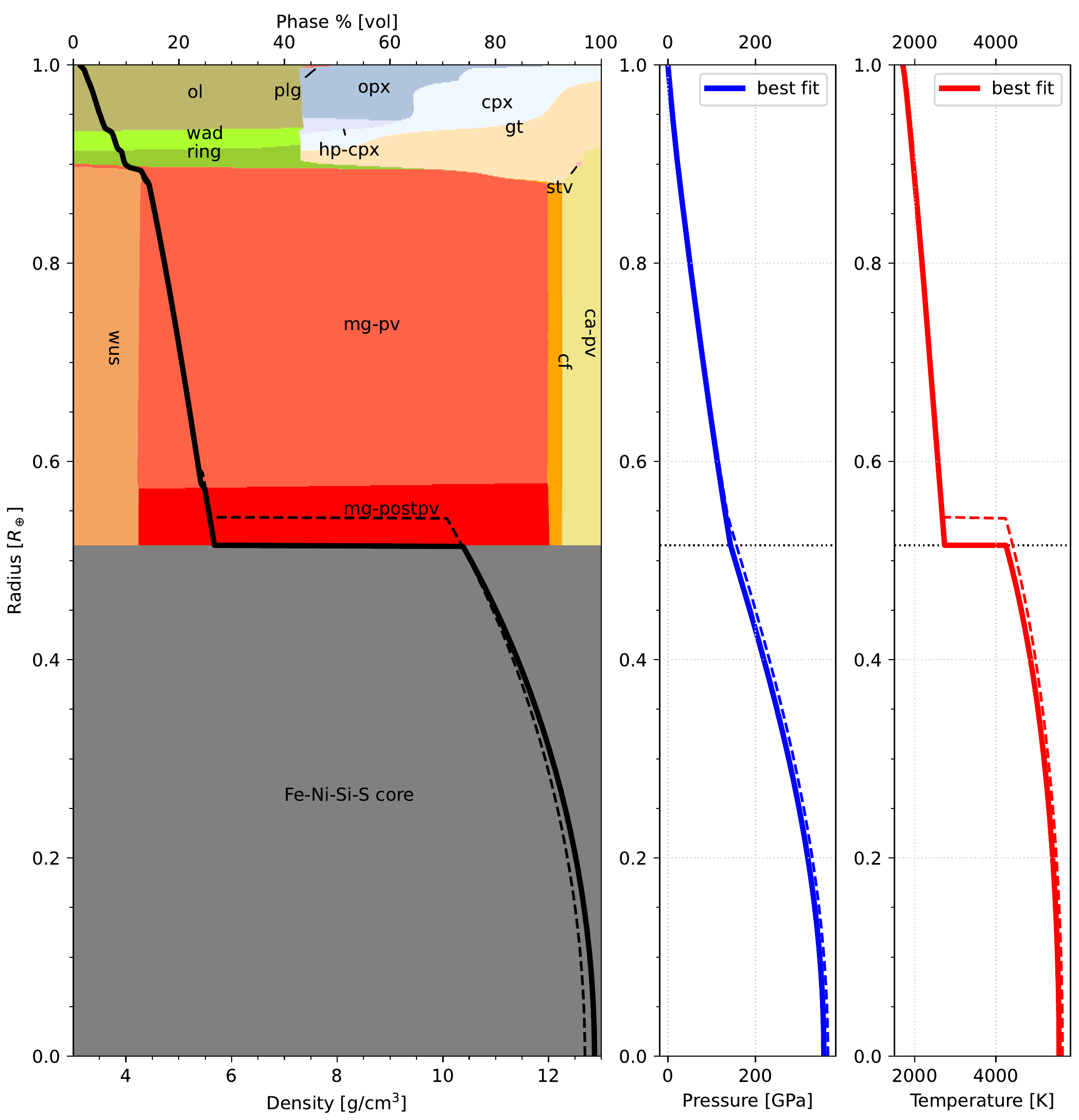}
	\caption{The best-fit interior mineralogy and structure of Kepler-21 exo-Earth (as an example). The solid black, blue and red curves are respectively the density, pressure and temperature profiles of the planet. For reference, the corresponding profiles of the Earth are shown as dashed curves. The grey regime on the left panel represents the core, which is presumably composed of Fe, Ni, Si, and S (the sum of which is normalised to 1). The temperature jumps at the CMBs are introduced by following \citet{Noack2020}, resulting in 1513 K for K21-exoE and 1537 K for Earth. The abbreviations for mantle mineralogy stand for: ol - olivine, plg - plagioclase, opx - orthopyroxene, cpx - clinopyroxene, hp-cpx - high-pressure clinopyroxene, wad - wadsleyite, ring - ringwoodite, gt - garnet, stv - stishovite, wus - magnesiow\"ustite (ferropericlase), mg-pv - magnesium perovskite (bridgmanite), mg-postpv - magnesium postperovskite, cf - calcium-ferrite structured phase, and ca-pv - calcium perovskite. For other details of the model see text.}
	\label{fig:K21_interior}
\end{figure*}

Using Kepler-21 exo-Earth (K21-exoE) as an example, we present its best-fit mineralogy and structure (in terms of self-consistent pressure, temperature and density profiles) in Fig. \ref{fig:K21_interior}. It shows that in the upper mantle, pyroxenes (orthopyroxene -- "opx", clinopyroxene -- "cpx" and high-pressure clinopyroxene -- "hp-cpx") are relatively enriched over olivine ("ol"), being consistent with the planet's relatively low Mg/Si ratio (the left panel of Fig. \ref{fig:keyR}). The lower mantle (starting from the density jump at $\sim$ 0.9 $R_{\oplus}$) is dominated by magnesium (post-)perovskite ("mg-pv" and "mg-postpv") and is similar to the Earth's lower mantle composition \citep{Palme2014b}. By comparing its density profile with the Earth's (in both cases a (plausible) inner solid core has been ignored; Fig. \ref{fig:K21_interior}), K21-exoE has a best-fit core slightly smaller than that of Earth. Please note that the best-fit result is obtained at the mean values of the first-order major oxides and core mass fraction (Table \ref{tab:interior}) and at a radius of 1 $R_{\oplus}$ (with its self-consistent mass, returned together with the mineralogy, equivalent to 1 $M_{\oplus}$ as well). The uncertainties of the best-fit results for each mineral in the mantle, the radius fraction of the core as well as other structural profiles (Fig. \ref{fig:mineral_uncertainty.pdf}) are obtained at the 16\% and 84\% quartiles of a population of the mineralogy and structure analyses repeated from a random draw of the estimates of the mantle and core compositions as well as core mass fractions (Table \ref{tab:interior}). 

The best-fit results for all planets, as shown in Figs. \ref{fig:mean_minerals_exoE_all_1} and \ref{fig:mean_minerals_exoE_all_2}, reveal that most of these exo-Earths are Earth-like in both the mineralogy and structure. The notable exceptions are K10-exoE and K37-exoE. The former has a negligible core -- as shown in Fig. \ref{fig:cmf} as well -- and a gigantic mantle, whereas the latter has a core that is comparably smaller than all cases other than K10-exoE. Further, both have a deep mantle dominated by the high-pressure, "mg-postpv" phase -- compared to the "mg-pv" phase (bridgemanite) for other cases. For the uppermost mantle (above $\sim$ 0.95 $R_{\oplus}$), it is dominated by olivine over pyroxene assemblages (also similar to that of Earth; \citealt{Palme2014b}) for all exo-Earth cases, except K2-277-exoE (and to a lesser degree, K21-exoE as well). Particularly, K2-277-exoE is also the only planet with a mean value of Mg/Si < 1 (Fig. \ref{fig:keyR}). This highlights the significance of Mg/Si in determining the mantle mineralogy, as investigated in \cite{Hinkel2018} and \citet{Spaargaren2020} as well. The intermediate range between upper and lower mantle ($\sim$ 0.9-0.95 $R_{\oplus}$; i.e. the mantle transition zone) is dominated by wadsleyite ("wad") and ringwoodite ("ring") (also similar to Earth's scenario), except for K10-exoE, K37-exoE, K68-exoE, and K2-277-exoE. Considering that both wadsleyite and ringwoodite can store water in their crystal structures by about one order of magnitude higher than any other minerals including olivine, pyroxene and perovskite \citep{Bercovici2003, Pearson2014, Fei2017}, these exceptional planets may also be the ones among the sample with the least water-storage capacity in its interior, although this assessment needs to be exercised with caution considering the yet-large uncertainty in the modelled mantle mineralogy. 

Broadly speaking and by considering the uncertainty level of such an analysis (as assessed for K21-exoE as an example), all of these cases (except K10-exoE and K37-exoE) share \textit{both} the internal structure \textit{and} mineralogy of a broadly Earth-like planet.

\section{DISCUSSION}
\label{sec:disc}
\subsection{The effect of varying devolatilisation scales on the interiors}
\label{sec:variance}
Planet formation is a complex process. A variety of outcomes for the bulk composition of a rocky planet
may result from composition-, location-, and time-scale-dependent differences in various devolatilisation processes \citep[W19b;][]{Dorn2019, Harrison2021}. Our empirical understanding of the devolatilisation from the protosolar to terrestrial abundances cannot be a true reflection of the devolatilisation that occurred in other planetary system, even for "exo-Earths" by our definition. However, we suppose that the discrepancy is not dramatic concerning the similar Earth-like composition for rocky exoplanets as revealed by the abundance measurements on polluted white dwarfs \citep{Doyle2019}. 

\begin{figure*}
 \includegraphics[trim=0cm 0cm 0cm 0cm, scale=0.65,angle=0]{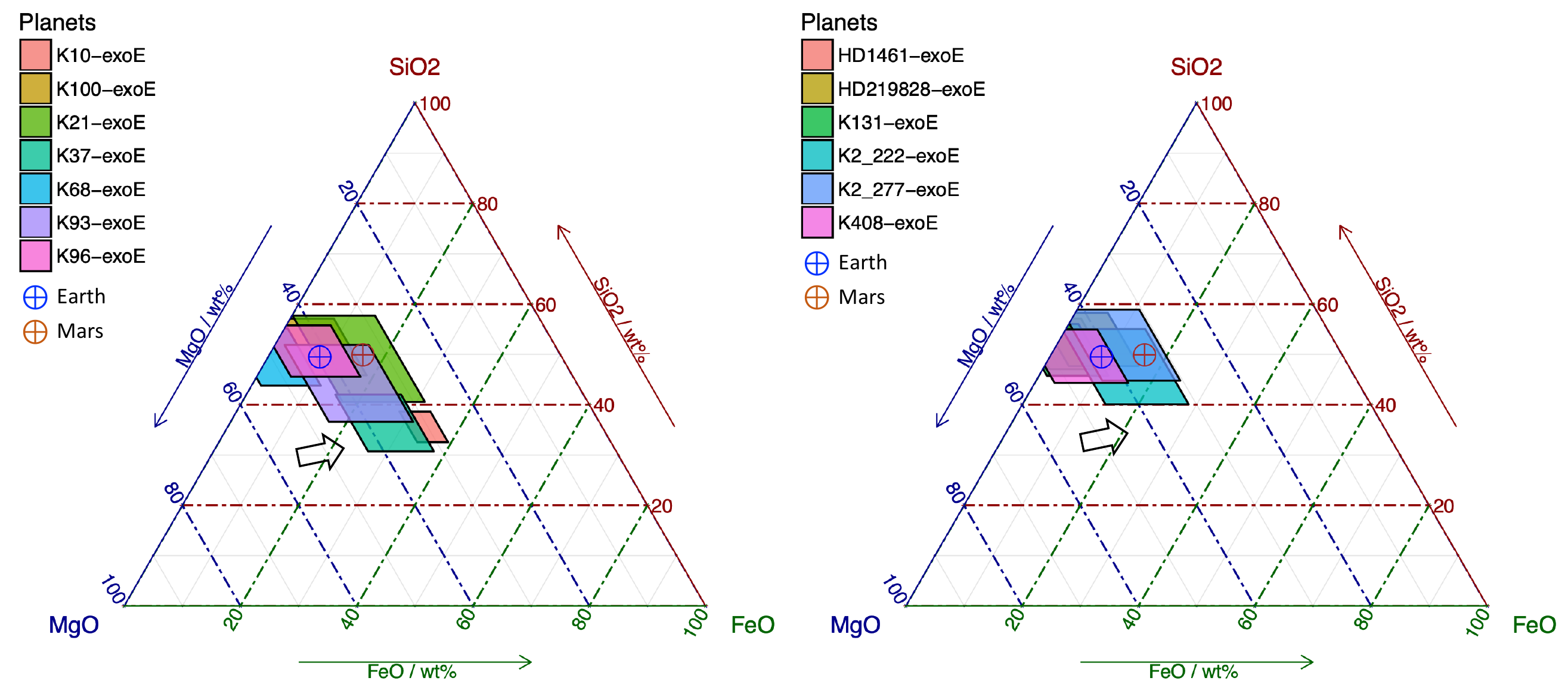}
	\caption{Similar to Fig. \ref{fig:ternary}, but constrained with the 3$\sigma$ upper limit of the Sun-to-Earth devolatilisation pattern (W19a). The open arrow indicates the overall direction that the mantle composition shifts towards in the parameter space of the ternary diagram. For details see Sect. \ref{sec:variance}.}
	\label{fig:ternary_ul}
\end{figure*}

\begin{figure*}
	\includegraphics[trim=0cm 0cm 0cm 0cm, scale=0.65,angle=0]{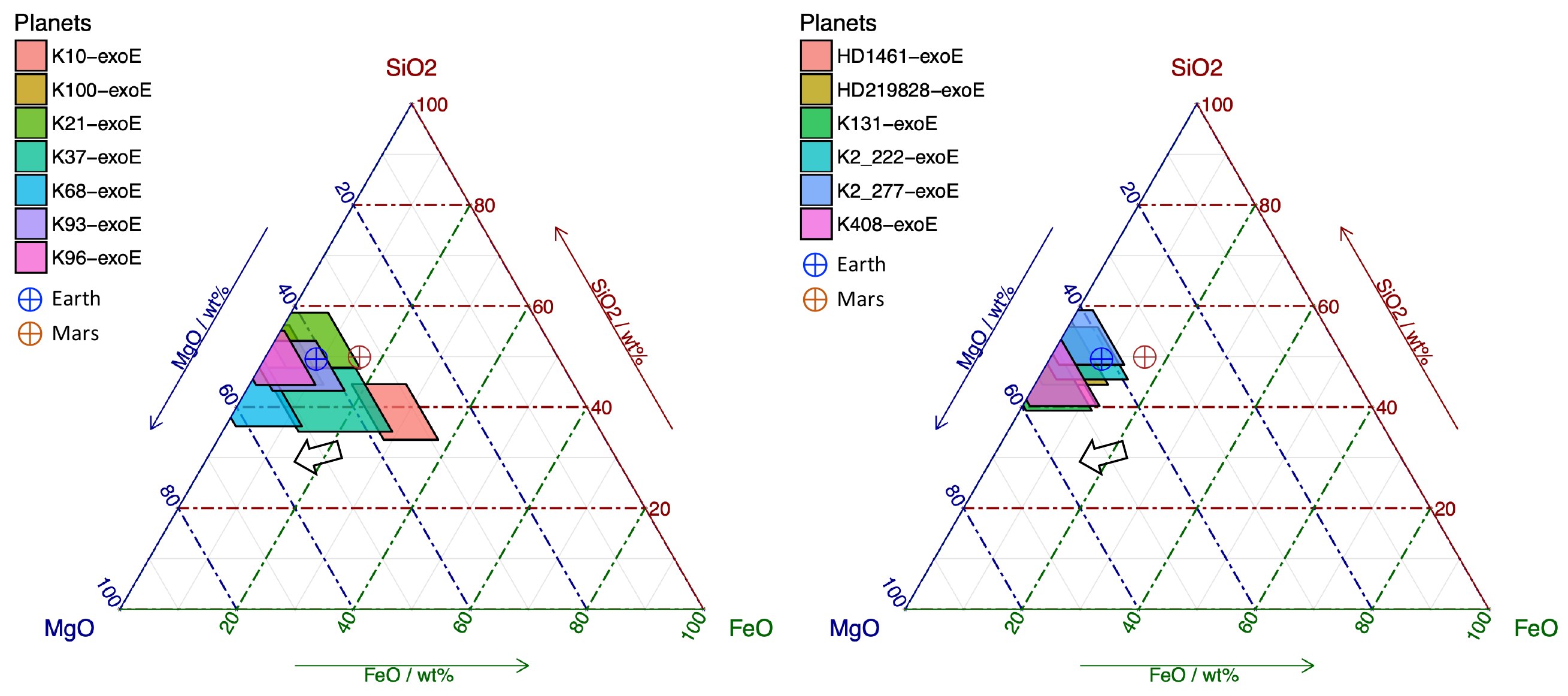}
	\caption{Similar to Fig. \ref{fig:ternary}, but constrained with the 3$\sigma$ lower limit of the Sun-to-Earth devolatilisation pattern (W19a). The open arrow indicates the overall direction that the mantle composition shifts towards in the parameter space of the ternary diagram. For details see Sect. \ref{sec:variance}.}
	\label{fig:ternary_ll}
\end{figure*}

To explore the effect of varying the devolatilisation scales on the interiors of exo-Earths, we apply the upper and lower limits of the 3$\sigma$ range of the adopted devolatilisation model to the abundances of these host stars. As a result,  we obtain two alternative sets of planetary bulk compositions, corresponding to the "less-depleted" (Fig. \ref{fig:ternary_ul}) and "more-depleted" (Fig. \ref{fig:ternary_ll}) scenarios, respectively. In the "less-depleted" scenario (Fig. \ref{fig:ternary_ul}), the mantle compositions (normalised by MgO + SiO$_2$ + FeO = 100 wt\%) of these exo-Earths shift towards the direction where SiO$_2$ and FeO are more enriched (relative to the scenario under the standard devolatilisation model -- Fig. \ref{fig:ternary}). In such a case, oxygen -- acting as a critical element to the mantle composition estimate -- is much less depleted, resulting in a much higher mantle oxidation state and thus more Si and Fe in the planet being oxidised. Likewise, in the "more-depleted" scenario (Fig. \ref{fig:ternary_ll}), the normalised mantle compositions shift towards the direction where SiO$_2$ and FeO are more depleted (relative to the scenario under the standard devolatilisation model -- Fig. \ref{fig:ternary}). In this case oxygen is much more depleted, thus resulting in a reduced mantle and an increased amount of Si as well as Fe partitioned into the core. 

The effect of such varied devolatilisation scales is also reflected onto the modelled core mass fractions (Fig. \ref{fig:cmf}). In the "less-depleted" scenario (shown as open diamonds withe error bars in Fig. \ref{fig:cmf}), the core mass fractions are systematically smaller than those under the "standard" scenario (i.e. filled circles with error bars in Fig. \ref{fig:cmf}), because the mantle is more oxidised with a reduced fractionation of metallic Fe and its alloy elements (Ni, Si, and S) into the core. Similarly, in the "more-depleted" scenario (shown as open circles with error bars in Fig. \ref{fig:cmf}), the core mass fractions are systematically larger than those under the "standard" scenario. In such a case, the mantle is more reduced and a larger fraction of Fe and its alloy elements is partitioned into the core. It is also noteworthy that the limits of our calculated CMFs (0--46\%) under the different scenarios are statistically broader than those constrained by the unaltered stellar compositions in the literature -- e.g. 20--46\% \citep{Plotnykov2020} and 21--41\% \citep{Schulze2021} -- and narrower than those constrained purely by mass and radius measurements -- e.g. 1--92\% \citep{Plotnykov2020} and 0--73\% \citep{Schulze2021} for (potentially) rocky planets. These discrepancies are fundamentally attributed to i) the difference in the available oxygen budget relative to other rock-forming elements in a planet (W19b) and ii) the inherent degeneracy in constraining planetary interiors with only mass and radius measurements \citep{Dorn2015}. For a detailed discussion of model and observational uncertainties on the determination of planetary interiors, we refer to \cite{Otegi2020}.

\begin{figure*}
	\includegraphics[trim=0cm 0cm 0cm 0cm, scale=1.0,angle=0]{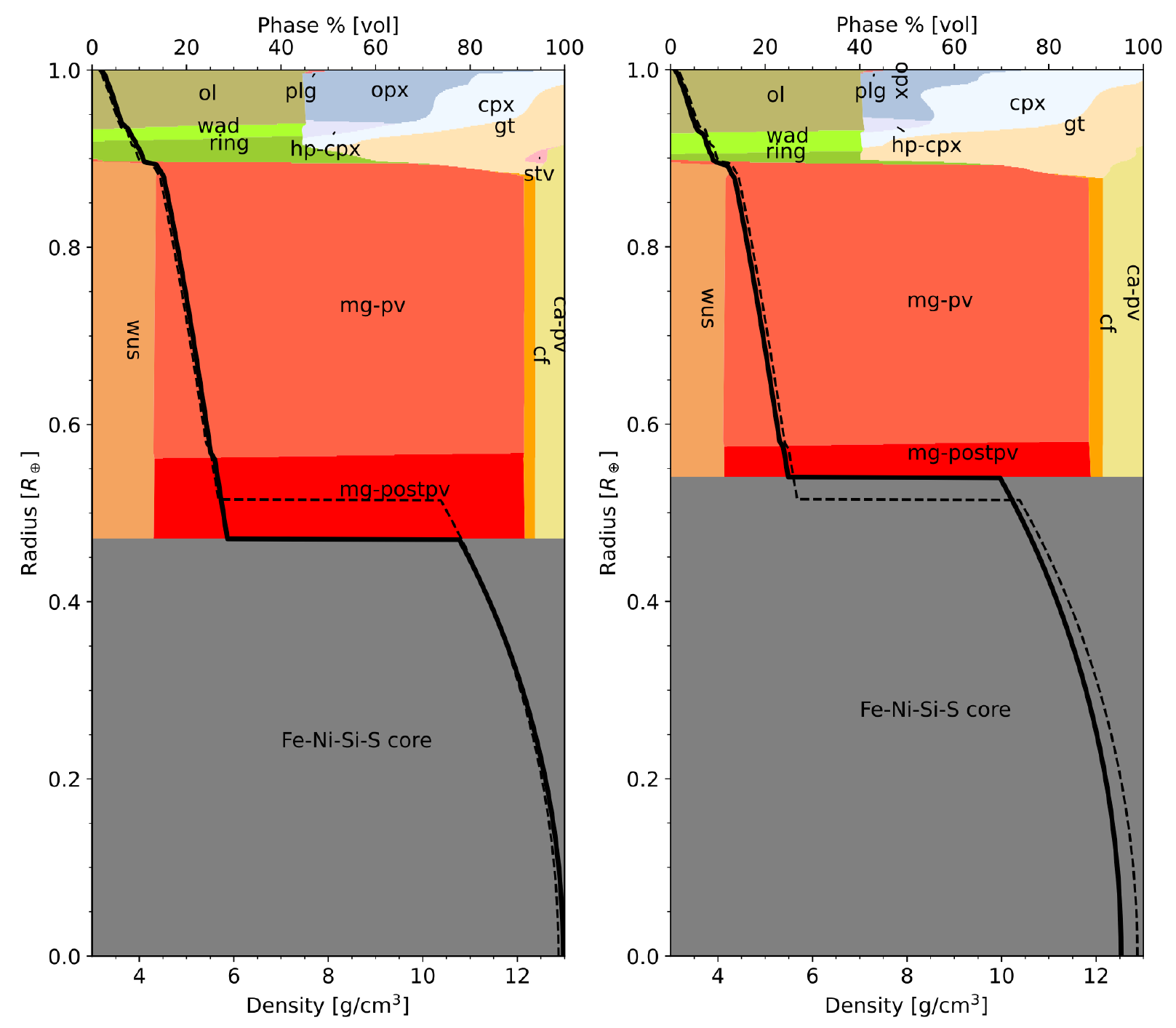}
	\caption{The modelled mineralogies and structures of Kepler-21 exo-Earth under two different devolatilisation scenarios: the 3$\sigma$ upper limit (left) and lower limit (right) of the Sun-to-Earth devolatilisation pattern (W19a). For comparison, the density profile for the planet under the "standard" scenario (Fig. \ref{fig:K21_interior}) is replotted as a dashed curve in each panel.}
	\label{fig:K21_interior_lmts}
\end{figure*}

Such a variance in the devolatilisation scales also induces a change to the modelled internal structure and mineralogy (Fig. \ref{fig:K21_interior_lmts}; K21-exoE is taken for example, in comparison with Fig. \ref{fig:K21_interior}). In the "less-depleted" scenario (left panel), the modelled core-mantle boundary (in line with the horizontal part of the solid black curve) is relatively deeper than that modelled under the "standard" scenario (dashed black curve), while it is opposite in the "more-depleted" scenario. For the mantle mineralogies as modelled under various scenarios, there are no significant differences, except for the minerals (e.g. ringwoodite -- "ring", wadsleyite -- "wad", and stishovite -- "stv") in the mantle transition zone and the high-pressure "mg-postpv" phase just above the core-mantle boundary. Overall, the variability of the devolatilisation scaling factors seems to influence more the internal structure than the mantle mineralogy. As explained in Sect. \ref{sec:ratios}, Mg/Si plays a critical role in modulating the mantle mineralogy, whereas this ratio is only \textit{negligibly} altered by devolatilisation (W19a). On the other hand, the fractionation of Fe between core and mantle is directly related to the mantle oxidation state \citep{Righter2003} and oxygen is a volatile (sensitive to devolatilisation); consequently, the internal structure may be more affected when the devolatilisation scale varies.  

It is noteworthy that our exploration of the effect of varied devolatilisation scales may not be applicable to peculiar planets such as super-Mercuries \citep[e.g.][]{Adibekyan2021}, nor to any type of planets around M stars such as the TRAPPIST-1 system \citep{Gillon2017} and Proxima b \citep{Anglada-Escude2016}. This is a consequence from potentially dramatic differences in planet formation histories and/or in stellar properties (e.g. XUV fluxes). To take into account these factors, a comprehensive investigation of nebular condensation \cite[e.g.][]{Wang2020a}, disc evolution \citep[e.g.][]{Bergner2020}, hydrodynamic escape \citep[e.g.][]{Benedikt2020}, accretionary dynamics \citep[e.g.][]{Emsenhuber2021}, and impacts \citep{Helffrich2019} is warranted. Before such a comprehensive investigation sheds more light on devolatilisation (in a quantitative manner), however, the inclusion of such an empirical model is a first-order but integral part of the efforts in reducing the modelling degeneracies of interiors of terrestrial-type exoplanets. 

\subsection{The effect of varying mass and radius (within the terrestrial regime) on the interiors}
\label{sec:MR}
\begin{figure*}
	\includegraphics[trim=0cm 0cm 0cm 0cm, scale=1.0,angle=0]{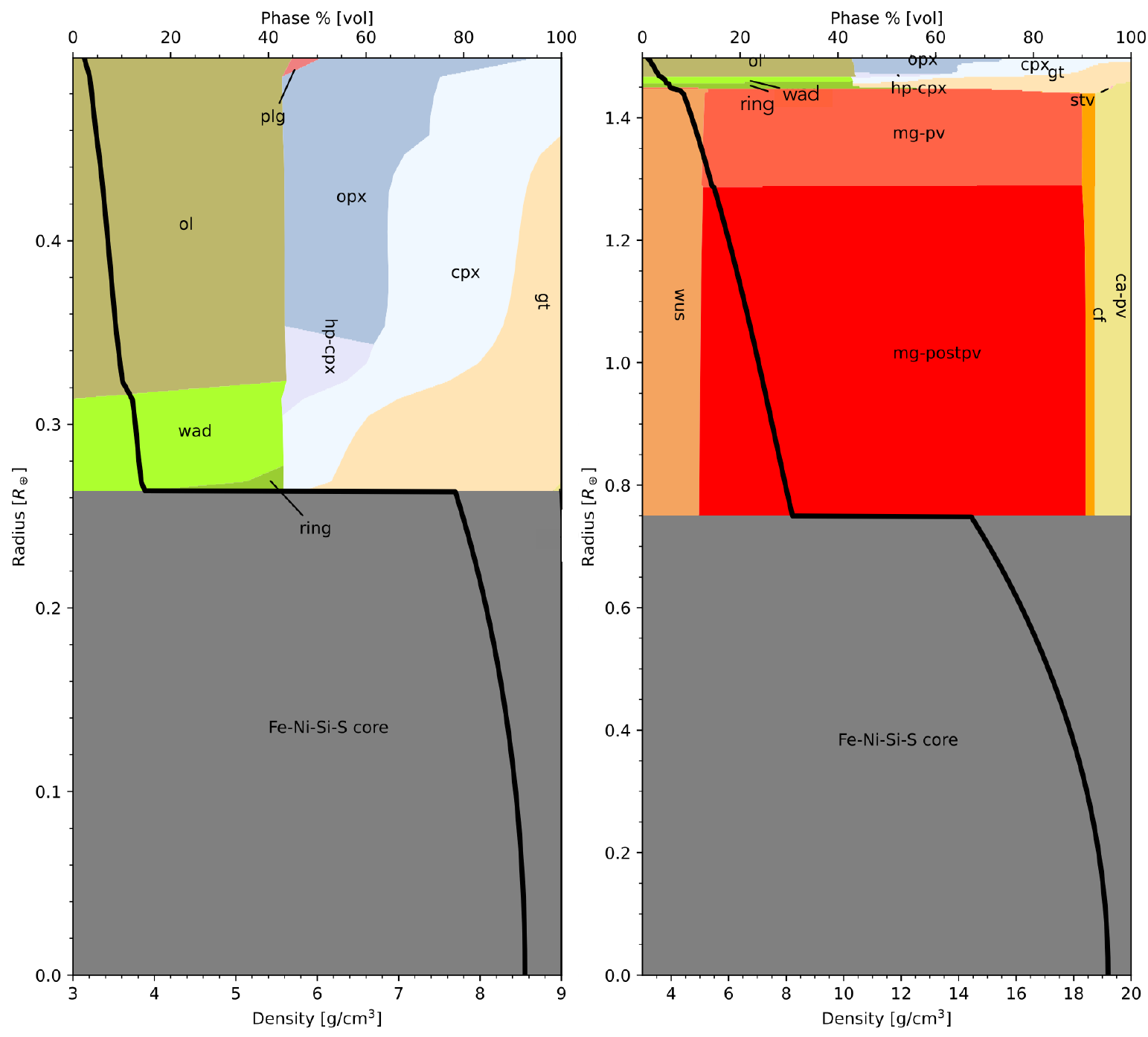}
	\caption{The modelled mineralogies and structures of Kepler-21 exo-Earth under the standard devolatilisation scenario but with different assumptions of planetary size: 0.5$R_{\oplus}$ (left) and 1.5$R_{\oplus}$ (right). The returned, self-consistent masses are 0.1$M_{\oplus}$ and 4.3$M_{\oplus}$, respectively, enabling the planet to be within the terrestrial regime. Please note that the CMB temperature jumps are introduced by following \citet{Stixrude2014}, instead of \citet{Noack2020} (which is parameterised for rocky planets of $[0.8, 2] M_\oplus$), for these two cases. Please also note the differences in the scales of radius and density between the two panels.}
	\label{fig:K21_interior_MR}
\end{figure*}

As analysed in Sect. \ref{sec:mineral}, these hypothetical exo-Earths are assumed to have a radius of 1 $R_{\oplus}$ and a mass of 1 $M_{\oplus}$. In reality, habitable-zone terrestrial planets are unlikely to be an ideal Earth 2.0 and their mass and radius may vary. To test the effect of varying mass and radius (within the terrestrial regime) on the interiors, we assume two extreme cases for the size of a model terrestrial planet by referring to the definition of a rocky planet orbiting within the (empirical) habitable zone: 0.5 $R_{\oplus}$ and 1.5 $R_{\oplus}$ for the LIFE targets \citep{Quanz2021}. The mass is not predefined but computed together with the interiors by keeping the planet to be in the terrestrial regime. The modelling results are shown in Fig. \ref{fig:K21_interior_MR}. 

For a model terrestrial planet of 0.5 $R_{\oplus}$ orbiting in the habitable zone of Kepler-21, its mantle mineralogy and self-consistent mass ($\sim$ 0.1 $M_{\oplus}$) approximately resemble those of Mars \citep{Yoshizaki2020}. For a model terrestrial planet of 1.5 $R_{\oplus}$ orbiting in the habitable zone of Kepler-21, however, there is no exact analogue in our Solar System, since its self-consistent mass of $\sim$ 4.3 $M_{\oplus}$ falls in the regime of a super-Earth and it has a lower mantle mineralogy distinctly dominated by the high-pressure "mg-postpv" phase, while its upper mantle mineralogy resembles that of Earth \citep{McDonough1995, Palme2014b}.

Please note that, both cases are conducted under the standard devolatilisation model. Namely, the same set of the first-order mantle and core compositions as well as core mass fraction of Kepler-21 exo-Earth (Table \ref{tab:interior}) are input for producing Figs. \ref{fig:K21_interior} and \ref{fig:K21_interior_MR}. In other words, the significant differences in mantle mineralogy between these scenarios shown in Figs. \ref{fig:K21_interior} and \ref{fig:K21_interior_MR} are dictated only by the size (and implicitly the mass) of the model planet. The trivial differences in core radius fraction are related to the difference in density compression between the smaller and larger (model) planets. 

\subsection{On the precision requirement of using (devolatilised) host stellar abundances to constrain exoplanet interiors}
\label{sec:precision}
In W19b, a precision level of $\lesssim$ 0.04 dex (differentially) or $\lesssim$ 10\% is proposed for the host stellar abundances to be used,  upon devolatilisation, to constrain and distinguish the interior compositions and structures of hypothetical terrestrial exoplanets. However, it is worth clarifying that this precision requirement has not taken into account the uncertainty in the reference solar abundance, which is typically at the level of $\sim$ 0.03 dex \citep{Asplund2009, Asplund2021} for the aforementioned major rock-forming elements. If the latter were taken into account, the precision requirement for the host stellar abundances should be as small as $\sim$ 0.025 dex (differentially) or $\sim$ 6\%, which has been achieved with high-precision spectroscopic analysis for bright Sun-like stars \citep[e.g. Paper I;][]{Spina2021} and binary stars \citep{Morel2018, Liu2018, Liu2021}. Apart from the concern with the precision, there are also nontrivial systematic differences between different reference solar abundances \citep[e.g.][]{Asplund2021, Adibekyan2021, Lodders2020, Wang2019a}. Therefore, one must clarify which solar abundances are applied for converting the stellar differential abundances to the absolute abundances, which are then used for further modelling of the properties of individual planets. For a detailed discussion of the systematic differences between stellar abundances determined with different techniques and their impacts on many aspects including modelling planetary properties, please refer to \cite{Hinkel2016}.

\subsection{Remaining limitations}
\label{sec:limitations}
For our interior structure model, we must emphasize that it is a rigid two-layer model - i.e. mantle + core, without the layers such as crust and water, nor that we have divided the core to an inner component and an outer one. Considering the currently typical uncertainties of 5-10\% and 10-20\%, respectively, for radius and mass measurements \citep{Otegi2020}, such a simplification is practical for interior modelling and has a trivial effect on our estimates of mantle composition and internal structure (in terms of core radius/mass fraction). The modelling of crust formation and composition is an active field and usually involves the modelling of partial melts, tectonics, and even weathering \citep{Rozel2017, Brugman2021, Hakim2021}, which are however beyond the scope of this paper. Nonetheless, our estimates of mantle mineralogy and core size provide essential information for future modelling of the crust owing to the fact that the crust is fundamentally an extraction product of the upper mantle through magmatic processes \citep{Rozel2017, Noack2012} and is also (ultimately) influenced by core formation \citep{Dyck2021}. 

A water layer (while without a crust layer) has been considered in other interior models \citep[e.g.][]{Tian2013, Noack2016, Dorn2017, Brugger2017, Unterborn2018, Acuna2021}. Based on numerical modelling, \cite{Shah2021} found that for a rocky planet with a mass of 0.1 -- 3 $M_{\oplus}$, the effect of an isolated surface water (i.e. oceans) on the radius is $\le$ 5\%, while the effect of a hydration case (i.e. water being chemically mixed with minerals) is $\le$ 2.5\% -- see also \cite{Vazan2022}. Such effects are still within the currently typical uncertainty for radius but may become more profound with the continuous improvement of the precisions of mass and radius measurements (e.g. up to 5\% and 3\%, respectively; \citealt{Stassun2017}). The effect of water inclusion on the mass of a terrestrial-type planet should be nonetheless negligible. However, water (regardless its amount) should be critically taken into account while extending a study on mineralogy and structure to that on internal dynamics \citep{Evans2014, Spaargaren2020} and/or crust formation \citep{Collins2020}.

Finally, we treat the core to be completely molten and homogeneously composed of Fe, Ni, Si and S, without differentiating a plausible inner solid portion made of pure Fe and Ni from an outer liquid portion that contains the light elements \citep{McDonough2003, Hirose2013, Wang2018}. This may inevitably underestimate the core mass and thus overestimate core radius should the planet's core be differentiated. However, we envisage that the under-/over-estimation should not be significant, concerning that Earth's inner solid core just accounts for 5\% of the mass of the core \citep{Yoder1995, McDonough2017} and that even a planet as small as Mars has been suggested to be in a molten state based on the most recent seismic data from the \textit{InSight} mission \citep{Stahler2021}. Conflicting views on liquid/solid cores for super-Earths have been presented \citep[e.g.][]{Valencia2006, Morard2011}, so do the models underpinning the two scenarios -- c.f., 'solid' \citep{Dorn2017, Brugger2017} vs. 'liquid' \citep{Unterborn2018, Lorenzo2018}. The phase of a planet's core should be not only related to the planet's mass and size but also its formation history and age \citep{Stevenson2008, Stixrude2014}. Before we know better such information collectively, this issue will remain open for further discussion. \\

\section{Summary and Conclusions}
\label{sec:conclusion}
Based on the detailed chemical compositions of planet-hosting stars obtained in Paper I and \cite{Liu2016}, this work extends the analysis of W19a,b and goes beyond the estimates of the first-order mantle oxide composition, core composition and core mass fraction to the detailed mineralogy (i.e. complex mineral phases that are often seen in a rocky planet like Earth) and interior structure (in terms of not only core mass fraction, but also core radius fraction and self-consistent density, pressure and temperature profiles). We have also assessed the uncertainties of the detailed interior estimates, which are contributed from the uncertainties associated with the host stellar abundances, the devolatilisation pattern, as well as the interior modelling degeneracy. Further, by respectively varying the devolatilisation scales and the planetary size, we test how such variabilities will affect the modelled interiors of terrestrial-type planets. 

We find that among the 13 model exo-Earths, 11 are broadly Earth-like in both interior composition and structure, whereas Kepler-10 and Kepler-37 exo-Earths (both with high (O-Mg-2Si)/Fe, implying a high planetary oxidation state) are predicted to have substantially smaller cores. On the effect of varying devolatilisation scaling factors on planetary interiors, we find that interior structure is more affected than the mantle mineralogy, probably because the fractionation of Fe between mantle and core is more sensitive to the planetary oxidation state -- thus determining the core size, whereas the mantle mineralogy is crucially modulated by Mg/Si \citep{Hinkel2018, Spaargaren2020} while this ratio is negligibly altered by devolatilisation (W19a). The further test of varying the planetary size (and implicitly mass, by keeping within the terrestrial regime) reveals the potential diversity of the mantle mineralogy of terrestrial planets even if they might have experienced an equivalent devolatilisation. We also recommend a precision level of $\lesssim$ 0.025 dex for the stellar differential abundances and a clarification of the reference solar abundances (and their uncertainties) in modelling planetary bulk composition and interiors. 

Our model is nonetheless limited by its rigid assumption of a two-layer structure -- i.e., mantle and core, with no crust (lithosphere) or water (potentially biosphere) considered yet, nor have we differentiated the core to be an inner component and an outer one based on their extent of solidification/crystallisation. However, such a simplification should have little impact on our modelling results of the static, mantle mineralogy and structure, while the model may be sophisticated further when the mass and radius measurements of real terrestrial planets become much more precise (e.g. up to 5\% and 3\%; \citealt{Stassun2017,Rauer2014}). 

Exercised with caution, such an analysis (with a yet-large uncertainty) nevertheless offers an insight, in terms what we can learn already with the available data (essentially host stellar composition as well as planetary mass and radius) and with the bulk/interior models (with sensibly simplifications) that we can build, into the detailed properties of habitable-zone, terrestrial-type exoplanets, thus providing guidance for the target selections for future missions, such as PLATO \citep{Rauer2014, Nascimbeni2022}, Ariel \citep{Turrini2021}, and LIFE \citep{Quanz2021,Quanz2022}.

\section*{Acknowledgements}
We thank the reviewers, particularly Lena Noack, for their helpful comments, which have greatly improved the quality of the manuscript.
This work has been carried out within the framework of the National Centre of Competence in Research PlanetS supported by the Swiss National Science Foundation (SNSF). H.S.W and S.P.Q acknowledge the financial support of the SNSF. FL acknowledges the support of the Australian Research Council through Future Fellowship grant FT180100194. S.J.M. thanks the Research Centre for Astronomy and Earth Sciences (Budapest, Hungary) for support.

\section*{Data availability}
The spectral data underlying this article are available in Keck Observatory Archive at \url{https://koa.ipac.caltech.edu/cgi-bin/KOA/nph-KOAlogin}. They can be accessed with Keck Program ID: Z148 (Semester: 2016B, PI: Yong) and Z279 (Semester: 2018A, PI: Yong). Other data underlying this article are available in the article or in the specified references. 


\bibliographystyle{mnras}
\bibliography{MyPapersBib} 

\appendix
\clearpage

\begin{landscape}
\section{Model planetary bulk compositions}

\FloatBarrier
\begin{table}[!htbp]
	\centering
	\caption{Model bulk compositions (normalised to Al=100) of hypothetical habitable-zone terrestrial exoplanets ("exoE"), as devolatilised$^a$ from their host stellar abundances$^b$. \\ The reported uncertainties are 1$\sigma$$^c$.}
	\begin{tabular}{l lllll lllll} 
		\toprule
		& C                   & O                   & S                   &Na                   &Si                   &Mg                   &Fe                   &Ni                   &Ca                   &Al                  \\
		\hline
       K10-exoE    &          41$^{+          10}_{-           8}$    &
3871$^{+         445}_{-         403}$    &
31$^{+           4}_{-           4}$    &
12$^{+           1}_{-           1}$    &
819$^{+          67}_{-          62}$    &
1048$^{+          92}_{-          84}$    &
679$^{+          72}_{-          65}$    &
38$^{+           4}_{-           4}$    &
66$^{+           5}_{-           5}$    &
100$^{+           7}_{-           7}$   \\
K21-exoE    &          57$^{+          15}_{-          12}$    &
4862$^{+         642}_{-         574}$    &
46$^{+           6}_{-           5}$    &
23$^{+           3}_{-           3}$    &
1490$^{+         128}_{-         117}$    &
1515$^{+         135}_{-         124}$    &
1225$^{+         138}_{-         124}$    &
66$^{+           8}_{-           7}$    &
110$^{+           9}_{-           8}$    &
100$^{+           7}_{-           7}$   \\
K37-exoE    &          30$^{+           7}_{-           6}$    &
2936$^{+         339}_{-         307}$    &
{-}    &
10$^{+           1}_{-           1}$    &
670$^{+          56}_{-          51}$    &
1013$^{+          85}_{-          78}$    &
617$^{+          65}_{-          59}$    &
34$^{+           4}_{-           3}$    &
59$^{+           5}_{-           4}$    &
100$^{+          28}_{-          22}$   \\
K68-exoE    &          22$^{+           5}_{-           4}$    &
1857$^{+         213}_{-         193}$    &
20$^{+           2}_{-           2}$    &
10$^{+           1}_{-           1}$    &
649$^{+          53}_{-          49}$    &
744$^{+          61}_{-          56}$    &
597$^{+          63}_{-          57}$    &
35$^{+           4}_{-           3}$    &
51$^{+           4}_{-           3}$    &
100$^{+          35}_{-          26}$   \\
K93-exoE    &          30$^{+           8}_{-           7}$    &
2889$^{+         332}_{-         301}$    &
28$^{+           3}_{-           3}$    &
12$^{+           1}_{-           1}$    &
776$^{+          64}_{-          59}$    &
996$^{+          83}_{-          76}$    &
716$^{+          75}_{-          68}$    &
41$^{+           4}_{-           4}$    &
65$^{+           6}_{-           5}$    &
100$^{+          12}_{-          11}$   \\
K96-exoE    &          36$^{+           9}_{-           7}$    &
3176$^{+         361}_{-         327}$    &
34$^{+           4}_{-           4}$    &
14$^{+           1}_{-           1}$    &
1000$^{+          81}_{-          75}$    &
1152$^{+          95}_{-          87}$    &
1026$^{+         107}_{-          97}$    &
54$^{+           6}_{-           5}$    &
86$^{+           6}_{-           6}$    &
100$^{+          10}_{-           9}$   \\
K100-exoE    &          42$^{+          10}_{-           9}$    &
2785$^{+         361}_{-         323}$    &
35$^{+           4}_{-           4}$    &
16$^{+           2}_{-           1}$    &
906$^{+          80}_{-          73}$    &
985$^{+          86}_{-          79}$    &
759$^{+          84}_{-          76}$    &
46$^{+           5}_{-           5}$    &
64$^{+           5}_{-           5}$    &
100$^{+          10}_{-           9}$   \\
K131-exoE    &          32$^{+           9}_{-           8}$    &
2459$^{+         291}_{-         263}$    &
28$^{+           3}_{-           3}$    &
13$^{+           1}_{-           1}$    &
861$^{+          71}_{-          65}$    &
963$^{+          82}_{-          75}$    &
847$^{+          91}_{-          82}$    &
49$^{+           5}_{-           5}$    &
71$^{+           6}_{-           5}$    &
100$^{+           8}_{-           7}$   \\
K2-222-exoE    &          54$^{+          22}_{-          17}$    &
3809$^{+         471}_{-         424}$    &
40$^{+           5}_{-           5}$    &
18$^{+           2}_{-           2}$    &
1125$^{+          96}_{-          88}$    &
1293$^{+         114}_{-         104}$    &
960$^{+         106}_{-          96}$    &
51$^{+           6}_{-           5}$    &
84$^{+           7}_{-           6}$    &
100$^{+          12}_{-          11}$   \\
K2-277-exoE    &          44$^{+          13}_{-          11}$    &
3012$^{+         394}_{-         353}$    &
37$^{+           5}_{-           4}$    &
14$^{+           2}_{-           1}$    &
966$^{+          86}_{-          78}$    &
958$^{+          87}_{-          79}$    &
851$^{+          96}_{-          86}$    &
51$^{+           6}_{-           5}$    &
70$^{+           7}_{-           6}$    &
100$^{+           8}_{-           7}$   \\
K408-exoE    &          44$^{+          11}_{-           9}$    &
3418$^{+         434}_{-         390}$    &
38$^{+           5}_{-           4}$    &
15$^{+           2}_{-           2}$    &
1125$^{+         101}_{-          92}$    &
1332$^{+         116}_{-         107}$    &
957$^{+         105}_{-          95}$    &
50$^{+           6}_{-           5}$    &
93$^{+           8}_{-           7}$    &
100$^{+           9}_{-           8}$   \\
HD1461-exoE    &          36$^{+           9}_{-           8}$    &
2426$^{+         293}_{-         264}$    &
30$^{+           4}_{-           3}$    &
17$^{+           2}_{-           2}$    &
867$^{+          72}_{-          66}$    &
928$^{+          78}_{-          72}$    &
762$^{+          82}_{-          74}$    &
49$^{+           5}_{-           5}$    &
59$^{+           5}_{-           4}$    &
100$^{+          10}_{-           9}$   \\
HD219828-exoE    &          33$^{+           9}_{-           7}$    &
2482$^{+         318}_{-         286}$    &
28$^{+           3}_{-           3}$    & 
16$^{+           2}_{-           1}$    &
840$^{+          71}_{-          65}$    &
868$^{+          77}_{-          70}$    &
730$^{+          81}_{-          73}$    &
44$^{+           5}_{-           4}$    &
60$^{+           5}_{-           5}$    &
100$^{+           9}_{-           9}$   \\

		\bottomrule
		\multicolumn{11}{p{16cm}}{\footnotesize$^a$ The model of devolatilisation: W19a -- \cite{Wang2019a}; the specific devolatilisation factors for these listed elements can also be found in Table 1 of W19b -- \cite{Wang2019b}.}\\
		\multicolumn{11}{p{16cm}}{\footnotesize$^b$ Sources of host stellar abundances: Paper I -- \cite{Liu2020} (Table 2), except for Kepler 10 \citep{Liu2016} (Table 2, derived with HET data)} \\ 					
		\multicolumn{11}{p{16cm}}{\footnotesize$^c$ The uncertainties are propagated from the 1$\sigma$ uncertainties in both host stellar abundances and in the devolatilisation model.} 
	\end{tabular}
	\label{tab:abu}
\end{table}
\FloatBarrier
\end{landscape}

\clearpage
\begin{landscape}
\section{Modelling details of planetary interiors}

\FloatBarrier
\begin{table}[!htbp]
	\centering
	\caption{Interior compositions and core mass fractions (with 1$\sigma$ uncertainties) of hypothetical exo-Earths (exoE) around the sample of stars, in comparison with Earth.}
	\begin{tabular}{l cccccc ccccc} 
		\toprule
	& \multicolumn{11}{c}{Mantle composition (wt\%)}\\
  &                Na$_2$O &                 CaO &                 MgO &         Al$_2$O$_3$ &             SiO$_2$ &                 FeO &                 NiO &              SO$_3$ &              CO$_2$ &            C & Metals \\
  \hline
       K10-exoE    &0.24$^{+0.04}_{-0.03}$    &2.41$^{+0.36}_{-0.24}$    &
27.4$^{+4.3}_{-2.6}$    &3.32$^{+0.47}_{-0.30}$    &31.8$^{+5.1}_{-2.9}$    &
29.02$^{+ 2.99}_{- 5.32}$    &1.66$^{+0.23}_{-0.38}$    &1.39$^{+0.26}_{-0.42}$
&0.68$^{+0.18}_{-0.17}$    &0.15$^{+0.05}_{-0.05}$    &
{-}   \\
K21-exoE    &0.41$^{+0.09}_{-0.09}$    &3.52$^{+0.62}_{-0.73}$    &
35.2$^{+6.2}_{-7.7}$    &2.92$^{+0.51}_{-0.60}$    &47.8$^{+5.3}_{-8.3}$    &
8.62$^{+ 6.89}_{- 6.58}$    &0.42$^{+0.47}_{-0.27}$    &0.50$^{+0.10}_{-0.10}$
&0.03$^{+0.01}_{-0.01}$    &0.38$^{+0.12}_{-0.10}$    &
0.15$^{+0.28}_{-0.13}$   \\
K37-exoE    &0.27$^{+0.06}_{-0.04}$    &2.87$^{+0.67}_{-0.45}$    &
35.0$^{+9.3}_{-5.5}$    &4.45$^{+1.58}_{-1.15}$    &34.4$^{+7.3}_{-5.5}$    &
21.01$^{+ 6.14}_{-11.56}$    &1.19$^{+0.42}_{-0.75}$    & {-}
&0.20$^{+0.06}_{-0.07}$    &0.27$^{+0.08}_{-0.07}$    &
0.02$^{+0.04}_{-0.02}$   \\
K68-exoE    &0.49$^{+0.07}_{-0.06}$    &4.32$^{+0.55}_{-0.47}$    &
45.6$^{+5.6}_{-5.0}$    &7.82$^{+2.73}_{-2.13}$    &41.0$^{+5.7}_{-6.9}$    &
0.22$^{+ 0.43}_{- 0.18}$    &0.03$^{+0.02}_{-0.01}$    &0.04$^{+0.01}_{-0.03}$
&{-}    &0.41$^{+0.10}_{-0.09}$    &
0.03$^{+0.15}_{-0.02}$   \\
K93-exoE    &0.35$^{+0.06}_{-0.07}$    &3.46$^{+0.61}_{-0.66}$    &
38.4$^{+6.5}_{-7.7}$    &4.82$^{+0.94}_{-0.96}$    &42.2$^{+5.0}_{-7.2}$    &
9.29$^{+ 7.87}_{- 7.10}$    &0.53$^{+0.51}_{-0.33}$    &0.61$^{+0.11}_{-0.12}$
&0.02$^{+0.01}_{-0.01}$    &0.34$^{+0.11}_{-0.09}$    &
0.02$^{+0.13}_{-0.02}$   \\
K96-exoE    &0.38$^{+0.05}_{-0.05}$    &4.31$^{+0.56}_{-0.59}$    &
41.5$^{+5.3}_{-5.8}$    &4.53$^{+0.68}_{-0.66}$    &45.7$^{+4.2}_{-5.5}$    &
2.47$^{+ 2.90}_{- 1.94}$    &0.12$^{+0.16}_{-0.07}$    &0.25$^{+0.05}_{-0.11}$
& {-}    &0.38$^{+0.10}_{-0.09}$    &
0.37$^{+0.52}_{-0.30}$   \\
K100-exoE    &0.49$^{+0.08}_{-0.08}$    &3.64$^{+0.54}_{-0.57}$    &
40.8$^{+5.6}_{-6.2}$    &5.20$^{+0.83}_{-0.83}$    &46.5$^{+4.4}_{-6.1}$    &
2.37$^{+ 3.20}_{- 1.89}$    &0.19$^{+0.20}_{-0.10}$    &0.30$^{+0.06}_{-0.13}$
& {-}   &0.51$^{+0.14}_{-0.12}$    &
0.03$^{+0.05}_{-0.02}$   \\
K131-exoE    &0.47$^{+0.07}_{-0.06}$    &4.56$^{+0.59}_{-0.54}$    &
44.7$^{+5.5}_{-5.2}$    &5.85$^{+0.77}_{-0.69}$    &43.5$^{+5.2}_{-6.2}$    &
0.39$^{+ 0.61}_{- 0.32}$    &0.04$^{+0.04}_{-0.02}$    &0.07$^{+0.02}_{-0.05}$
& {-}   &0.43$^{+0.14}_{-0.11}$    &
0.05$^{+0.37}_{-0.04}$   \\
K2-222-exoE    &0.40$^{+0.08}_{-0.08}$    &3.45$^{+0.57}_{-0.65}$    &
38.5$^{+5.9}_{-7.7}$    &3.72$^{+0.71}_{-0.73}$    &45.5$^{+4.8}_{-7.1}$    &
6.92$^{+ 6.38}_{- 5.37}$    &0.31$^{+0.43}_{-0.19}$    &0.51$^{+0.09}_{-0.13}$
&0.02$^{+0.01}_{-0.01}$    &0.46$^{+0.20}_{-0.16}$    &
0.19$^{+0.31}_{-0.17}$   \\
K2-277-exoE    &0.42$^{+0.08}_{-0.08}$    &3.72$^{+0.62}_{-0.67}$    &
36.8$^{+5.6}_{-6.9}$    &4.83$^{+0.74}_{-0.85}$    &48.7$^{+4.5}_{-6.7}$    &
4.25$^{+ 5.32}_{- 3.37}$    &0.31$^{+0.33}_{-0.17}$    &0.45$^{+0.09}_{-0.15}$
& {-}   &0.49$^{+0.16}_{-0.14}$    &
0.02$^{+0.14}_{-0.02}$   \\
K408-exoE    &0.39$^{+0.07}_{-0.06}$    &4.30$^{+0.63}_{-0.58}$    &
44.2$^{+6.1}_{-5.8}$    &4.19$^{+0.59}_{-0.59}$    &44.2$^{+4.8}_{-6.2}$    &
1.55$^{+ 1.89}_{- 1.19}$    &0.07$^{+0.10}_{-0.04}$    &0.17$^{+0.03}_{-0.08}$
& {-}   &0.44$^{+0.12}_{-0.10}$    &
0.45$^{+0.64}_{-0.36}$   \\
HD1461-exoE    &0.61$^{+0.10}_{-0.08}$    &3.88$^{+0.52}_{-0.46}$    &
43.8$^{+5.8}_{-5.1}$    &5.97$^{+0.89}_{-0.80}$    &44.7$^{+5.2}_{-6.5}$    &
0.33$^{+ 0.69}_{- 0.27}$    &0.05$^{+0.04}_{-0.02}$    &0.07$^{+0.02}_{-0.05}$
& {-}    &0.51$^{+0.14}_{-0.12}$    &
0.06$^{+0.08}_{-0.04}$   \\
HD219828-exoE    &0.57$^{+0.09}_{-0.09}$    &3.85$^{+0.57}_{-0.55}$    &
40.4$^{+5.6}_{-5.6}$    &5.86$^{+0.91}_{-0.85}$    &46.9$^{+4.5}_{-6.2}$    &
1.64$^{+ 2.26}_{- 1.37}$    &0.13$^{+0.14}_{-0.07}$    &0.20$^{+0.04}_{-0.11}$
& {-}    &0.45$^{+0.13}_{-0.11}$    &
0.03$^{+0.15}_{-0.02}$   \\
Earth$^a$ & 0.36 & 3.58 & 38.1 & 4.49 & 45.4 & 8.11 & {-} & {-} & {-} & {-} & {-}\\
\midrule
 & \multicolumn{8}{c}{Core composition (wt\%) and core mass fraction (CMF, wt\%)}\\
   & &        Fe &        Ni &        Si &         S & &       CMF\\
 \hline
K10-exoE & &90.4$^{+2.1}_{-3.1}$    & 5.5$^{+0.1}_{-0.2}$    &
0.2$^{+0.2}_{-0.1}$    & 3.9$^{+3.5}_{-2.0}$    & & 0.0$^{+ 9.4}_{- 0.0}$   \\
K21-exoE & &89.7$^{+1.7}_{-7.8}$    & 5.2$^{+0.3}_{-0.8}$    &
3.3$^{+3.1}_{-2.3}$    & 1.7$^{+0.5}_{-0.4}$    & &28.8$^{+ 6.5}_{-15.8}$   \\
K37-exoE & &93.4$^{+0.0}_{-1.1}$    & 5.7$^{+0.0}_{-0.5}$    &
0.9$^{+1.1}_{-0.7}$    & {-}    & &11.9$^{+15.0}_{-11.9}$   \\
K68-exoE & &80.3$^{+7.1}_{-6.3}$    & 5.0$^{+0.7}_{-0.7}$    &
13.2$^{+6.1}_{-6.5}$    & 1.5$^{+0.3}_{-0.2}$    & &38.7$^{+ 4.2}_{- 4.4}$   \\
K93-exoE & &90.6$^{+1.1}_{-5.8}$    & 5.5$^{+0.2}_{-0.7}$    &
2.1$^{+2.3}_{-1.5}$    & 1.8$^{+0.6}_{-0.5}$    & &26.9$^{+ 7.1}_{-14.2}$   \\
K96-exoE & &88.0$^{+4.9}_{-6.7}$    & 4.9$^{+0.7}_{-0.6}$    &
5.5$^{+4.2}_{-3.5}$    & 1.6$^{+0.3}_{-0.3}$    & &36.4$^{+ 4.4}_{- 6.7}$   \\
K100-exoE & &85.2$^{+7.0}_{-8.2}$    & 5.4$^{+0.4}_{-0.8}$    &
7.3$^{+5.2}_{-4.6}$    & 2.1$^{+0.4}_{-0.4}$    & &33.7$^{+ 5.0}_{- 7.7}$   \\
K131-exoE & &82.5$^{+7.1}_{-6.1}$    & 5.0$^{+0.7}_{-0.7}$    &
10.9$^{+5.5}_{-5.8}$    & 1.5$^{+0.3}_{-0.2}$    & &39.7$^{+ 4.4}_{- 4.7}$   \\
K2-222-exoE & &89.3$^{+2.0}_{-8.1}$    & 5.1$^{+0.4}_{-0.8}$    &
3.6$^{+3.2}_{-2.5}$    & 1.9$^{+0.5}_{-0.4}$    & &29.4$^{+ 5.7}_{-13.6}$   \\
K2-277-exoE & &87.8$^{+3.9}_{-8.3}$    & 5.4$^{+0.3}_{-0.7}$    &
4.8$^{+4.1}_{-3.1}$    & 2.0$^{+0.5}_{-0.5}$    & &33.6$^{+ 5.4}_{-10.8}$   \\
K408-exoE & &84.1$^{+8.5}_{-8.1}$    & 4.6$^{+0.8}_{-0.6}$    &
9.4$^{+6.1}_{-5.6}$    & 1.9$^{+0.4}_{-0.3}$    & &34.1$^{+ 4.7}_{- 5.9}$   \\
HD1461-exoE & &81.0$^{+7.7}_{-6.7}$    & 5.4$^{+0.6}_{-0.7}$    &
11.9$^{+6.2}_{-6.4}$    & 1.8$^{+0.3}_{-0.3}$    & &38.3$^{+ 4.5}_{- 4.8}$   \\
HD219828-exoE & &84.5$^{+8.1}_{-7.6}$    & 5.3$^{+0.5}_{-0.7}$    &
8.5$^{+5.4}_{-5.1}$    & 1.8$^{+0.3}_{-0.3}$    & &35.9$^{+ 4.9}_{- 6.5}$   \\
Earth$^b$ & & 87.5 & 5.3 & 5.2 & 1.9 && $32.5\pm0.3$\\
\bottomrule
\multicolumn{8}{p{15cm}}{\footnotesize $^a$ Refer to \citet{McDonough1995} and normalise the contents of the adopted six major oxides to 100 wt\%.}\\
\multicolumn{8}{p{15cm}}{\footnotesize $^b$ Refer to \citet{Wang2018} and normalise the contents of the adopted four elements in the core to 100 wt\%.}
\end{tabular}
\label{tab:interior}
\end{table}

\end{landscape}
\clearpage

\begin{figure*}
	\includegraphics[trim=0cm 0cm 0cm 0cm, scale=0.65,angle=0]{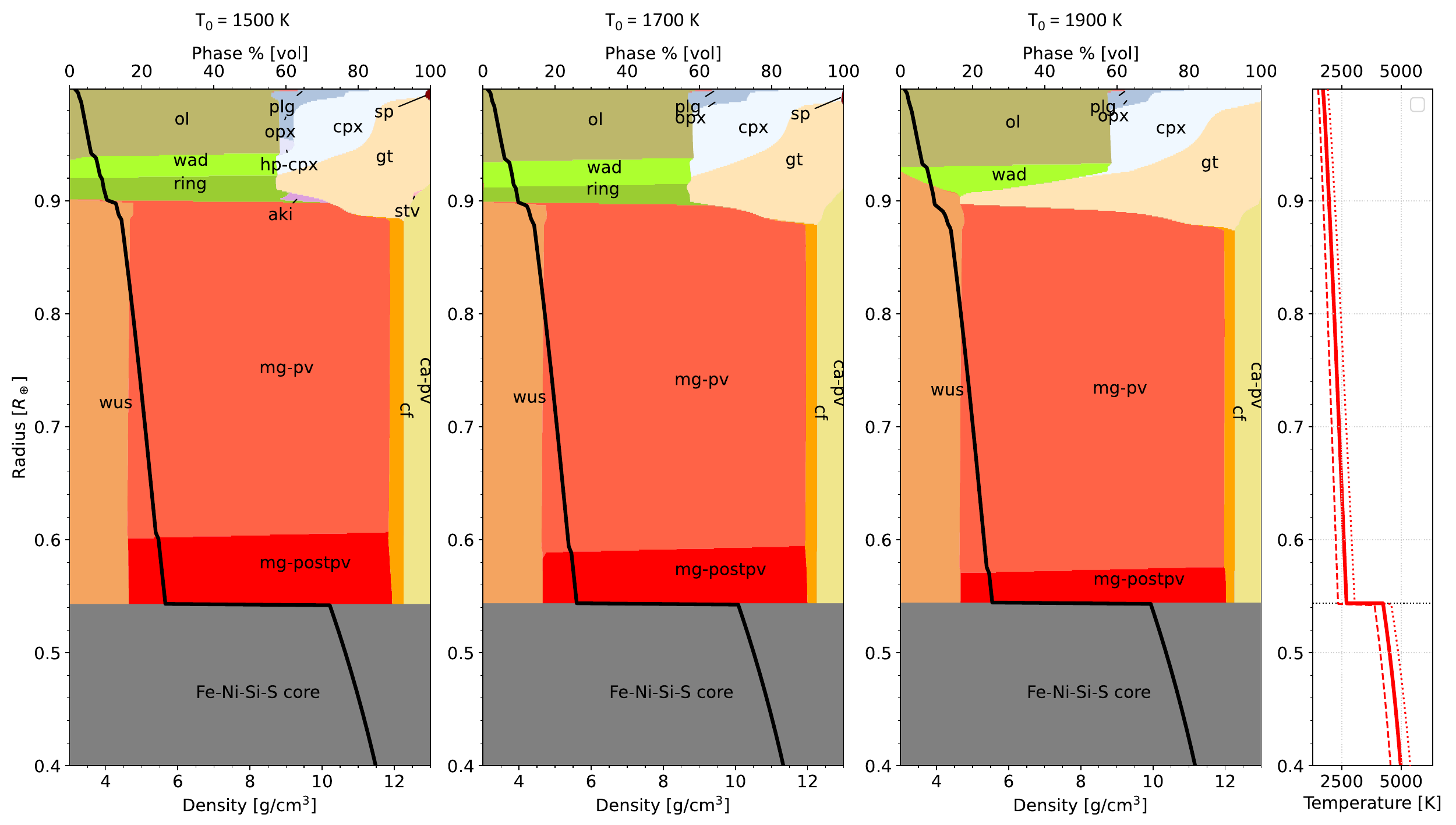}
	\caption{The comparison of the modelled mantle mineralogies in the case of Earth given different mantle potential temperatures ($T_0$): 1500 K, 1700 K, and 1900 K, which respectively correspond to the dashed, solid, and dotted curves in the rightmost panel for the temperature profiles. The core-mantle temperature jump is fixed at 1537 K for all cases. The core in each case is only partially shown to highlight the mantle mineralogy.}  
	\label{fig:mineral_Earth_varyTp.pdf}
\end{figure*}

\begin{figure*}
	\includegraphics[trim=0cm 0cm 0cm 0cm, scale=0.48,angle=0]{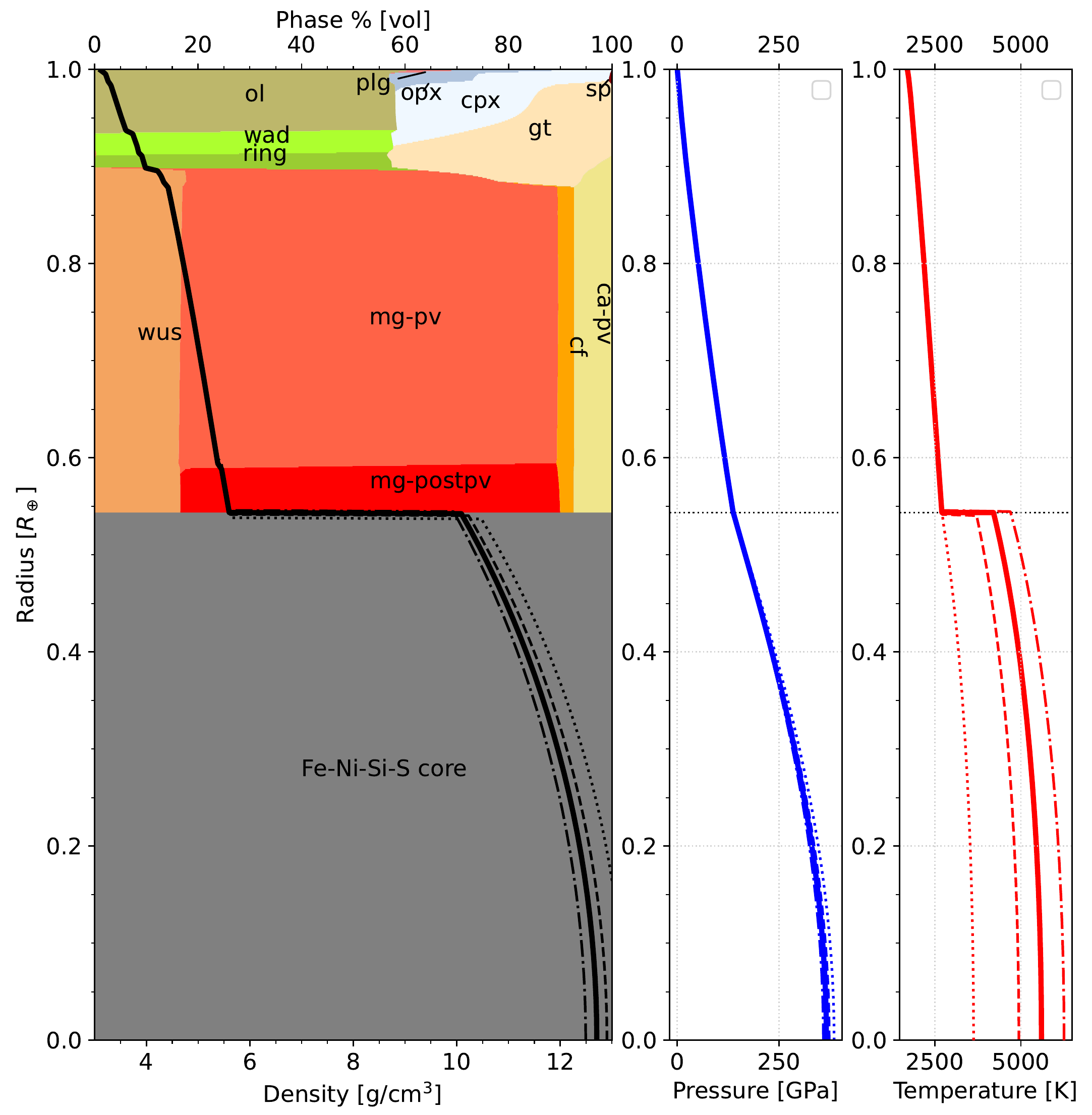}
	\caption{The comparison of the modelled interior profiles for density (black curves), pressure (blue curves), and temperature (red curves) in the case of Earth given different core-mantle temperature jumps: 0 K, 1000 K, 1500 K, and 2000 K, which respectively correspond to the dotted, dashed, solid, and dash-dotted curves of these profiles. The mantle potential temperature is fixed at 1700 K for all cases.}  
	\label{fig:Various_Delta_T_CMB.pdf}
\end{figure*}

\begin{figure*}
	\includegraphics[trim=0cm 0cm 0cm 0cm, scale=0.6,angle=0]{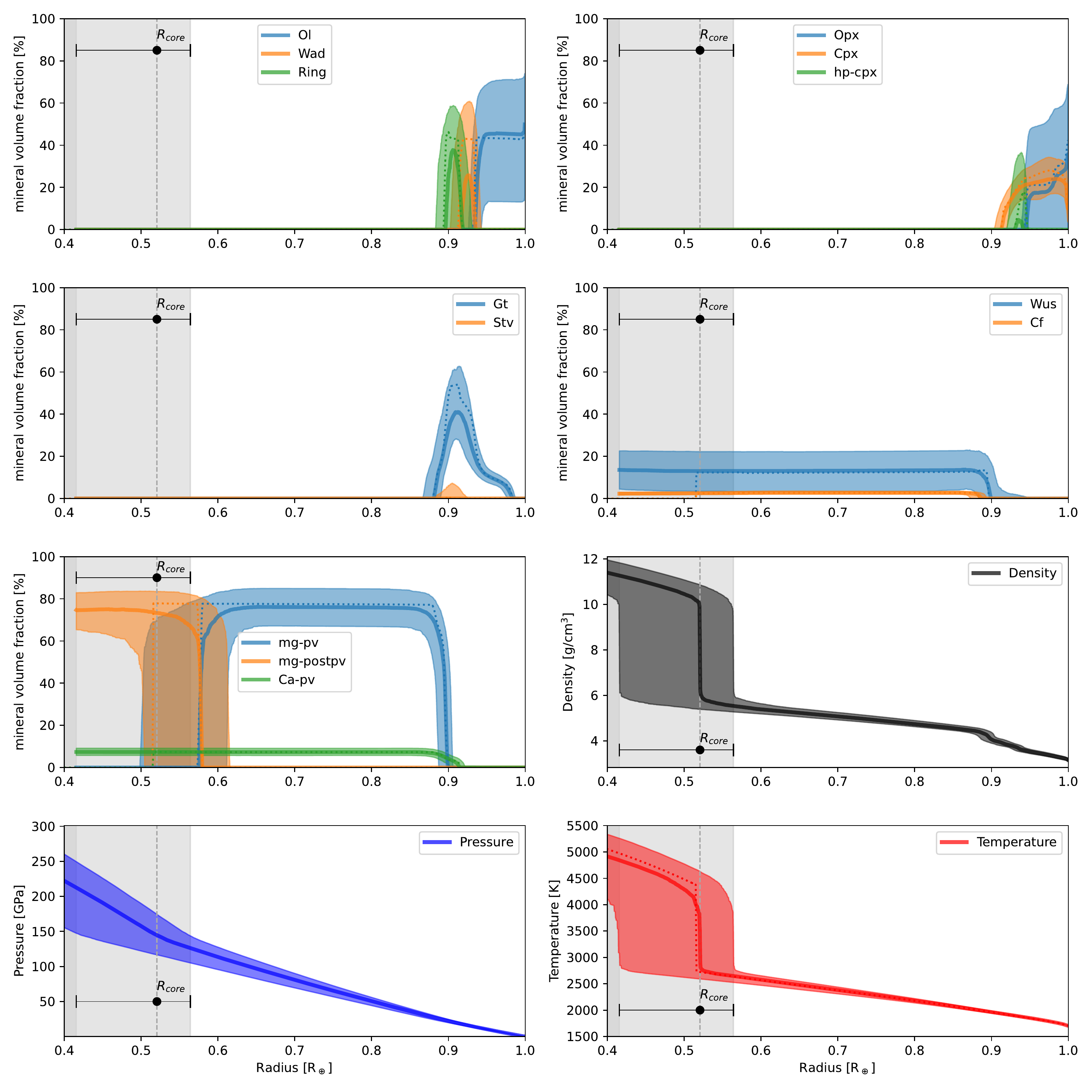}
	\caption{The uncertainty assessment of the modelled mineralogy and structure (taking K21-exoE as an example). The best-fit mineralogy/structure profiles (corresponding to Fig. \ref{fig:K21_interior}) are shown in dotted lines in each panel, with the 50th percentile and the range of 16th and 84th percentiles of the estimates shown in a solid line and a shadow area. The best-fit profiles of pressure and density are exactly coincident with their 50th percentiles. The vertical dashed line in each panel indicates the 50th percentile of the core size ($R_{\mathrm{core}}$), with its uncertainty shown in the light grey area. The illustration of the 50th percentile and the embraced range of [16th, 84th] percentiles of the lower mantle mineralogy has been cut off beyond the 1$\sigma$ lower limit of the radius.}  
	\label{fig:mineral_uncertainty.pdf}
\end{figure*}

\begin{figure*}
	\includegraphics[trim=0cm 0cm 0cm 0cm, scale=0.35,angle=0]{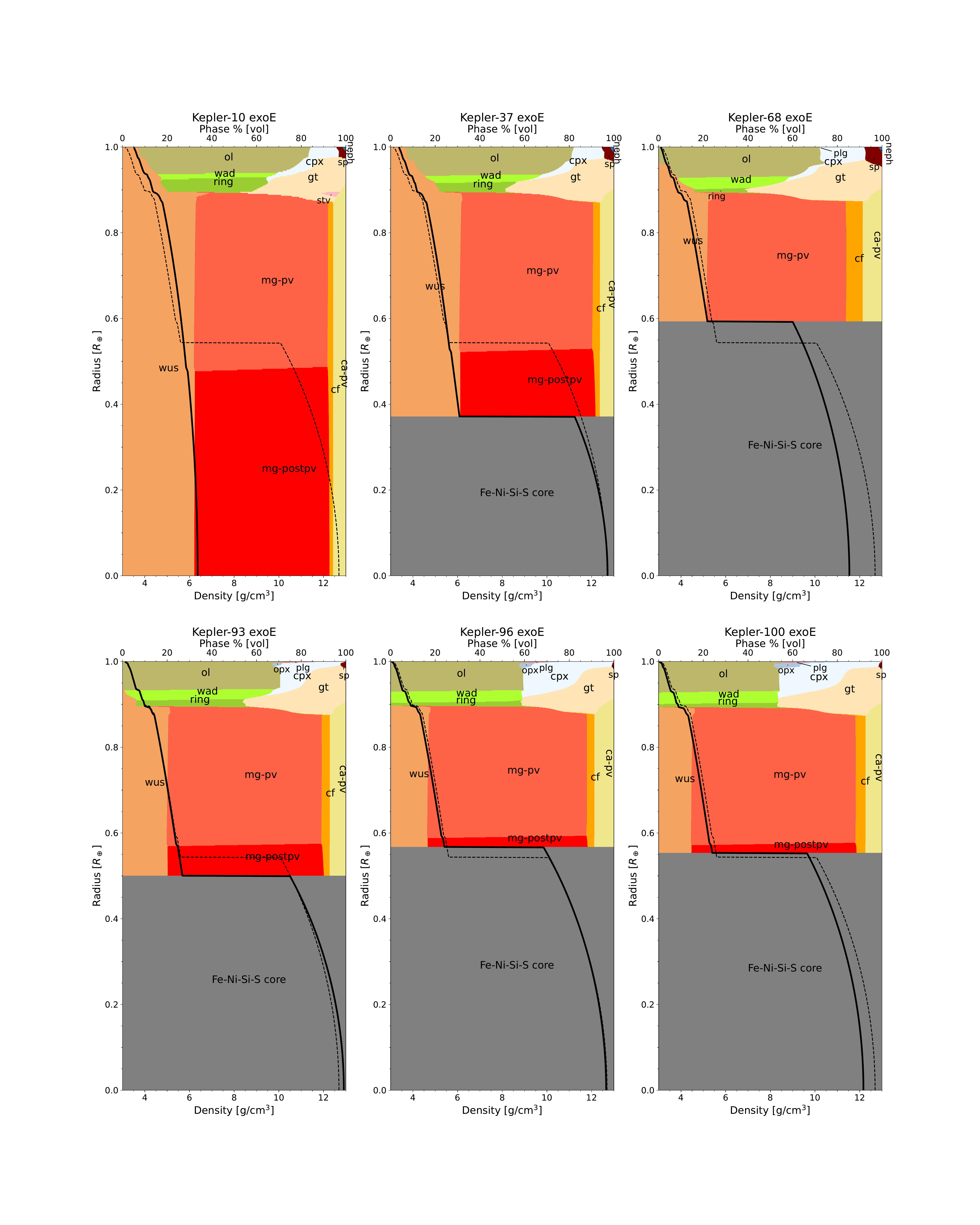}
	\caption{The best-fit mineralogies and structures of half of the sample of exo-Earths. The dashed curves indicate the Earth density profile. For the explanation of other details of these individual diagrams, please refer to Fig. \ref{fig:K21_interior}. The additional mineral name abbreviations -- sp and neph -- stand for spinel and nepheline, respectively.}
	\label{fig:mean_minerals_exoE_all_1}
\end{figure*}

\begin{figure*}
	\includegraphics[trim=0cm 0cm 0cm 0cm, scale=0.35,angle=0]{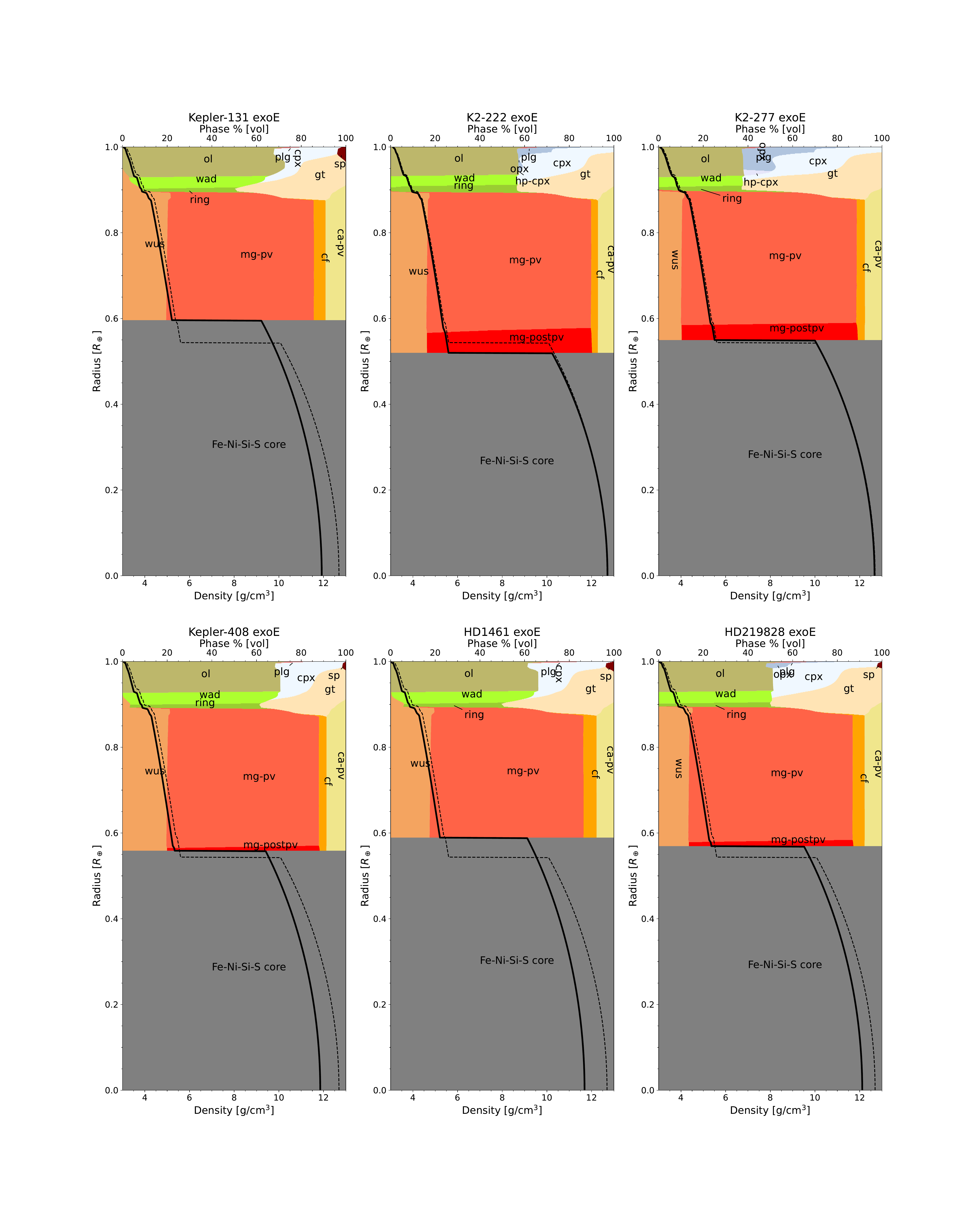}
	\caption{Similar to Fig. \ref{fig:mean_minerals_exoE_all_1} but for the other half of the sample.}
	\label{fig:mean_minerals_exoE_all_2}
\end{figure*}

\label{lastpage}
\end{document}